\documentclass[12pt]{article}
\usepackage{amsmath,amssymb,amsthm,amscd}
\usepackage{epsf,amsfonts,hyperref}
\usepackage{cite}
\usepackage{graphicx}

\input epsf.sty
\newcommand{\CA}{{\cal A}}
\newcommand{\CB}{{\cal B}}

\newcommand{\CG}{{\cal G}}

\newcommand{\CM}{{\cal M}}
\newcommand{\CN}{{\cal N}}
\newcommand{\CO}{{\cal O}}

\def\IZ{{\mathbb Z}}
\def\IR{{\mathbb R}}
\def\IC{{\mathbb C}}
\def\IP{{\mathbb P}}

\def\IS{{\mathbb S}}

\newcommand{\re}{{\rm e}}
\newcommand{\ri}{{\rm i}}
\newcommand{\rd}{{\rm d}}

\newcommand{\be}{\begin{equation}}
\newcommand{\ba}{\begin{aligned}}
\newcommand{\ea}{\end{aligned}}
\newcommand{\sectiono}[1]{\section{#1}\setcounter{equation}{0}}

\newcommand{\figref}[1]{Fig.~\protect\ref{#1}}

\newcommand{\beq}{\begin{equation}}
\newcommand{\eeq}{\end{equation}}
\newcommand{\bea}{\begin{eqnarray}}
\newcommand{\eea}{\end{eqnarray}}

\renewcommand{\and}{{\qquad {\rm and} \qquad}}

 \newcommand{\Tr}{{\,\rm Tr}\:}

\newcommand{\Res}{\mathop{\,\rm Res\,}}

\newcommand{\td}[1]{{\tilde{#1}}}

\renewcommand{\l}{\lambda}
\newcommand{\om}{\omega}

\renewcommand{\d}{{{\partial}}}

\newcommand{\Pint}{{\int\kern -1.em -\kern-.25em}}

\renewcommand{\l}{\lambda}

\newcommand{\acycle}{{\cal A}}
\newcommand{\bcycle}{{\cal B}}

\newcommand{\ee}[1]{{{\rm e}^{#1}}}

\newif\iffigs\figsfalse
\figstrue

\iffigs
  \input epsf
\else
\fi

\begin{document}

\begin{titlepage}
{}~
\hfill\vbox{
\hbox{IPhT-T08/164}
}\break

\vskip .6cm

\centerline{\Large \bf
A holomorphic and background independent partition function}
\vspace*{1.0ex}
\centerline{\Large \bf
for matrix models and topological strings}

\medskip

\vspace*{4.0ex}

\centerline{\large \rm Bertrand Eynard$^a$ and Marcos Mari\~no$^b$}

\vspace*{4.0ex}

\centerline{\rm ~$^a$  Institut de Physique Th\'eorique, CEA, IPhT}
\centerline{ F-91191 Gif-sur-Yvette, France}
\centerline{CNRS, URA 2306, F-91191 Gif-sur-Yvette, France }

\centerline{{\tt eynard@spht.saclay.cea.fr}}

\vspace*{1.8ex}

\centerline{ \rm ~$^b$Section de Math\'ematiques et D\'epartement de Physique Th\'eorique}
\centerline{ \rm Universit\'e de Gen\`eve, CH-1211 Gen\`eve, Switzerland}

\centerline{\tt
marcos.marino@unige.ch}

\vspace*{6ex}

\centerline{\bf Abstract}
\medskip
We study various properties of a nonperturbative partition function which can be associated to any spectral curve. When the spectral curve 
arises from a matrix model, this nonperturbative partition function is given by a sum of matrix integrals over all possible 
filling fractions, and includes all the multi-instanton 
corrections to the perturbative $1/N$ expansion. We show that the nonperturbative partition function, which is manifestly holomorphic, 
is also modular and background independent: it transforms as the partition function of a twisted 
fermion on the spectral curve. Therefore, modularity is restored by nonperturbative corrections. 
We also show that this nonperturbative partition function obeys the Hirota equation and provides a natural 
nonperturbative completion for topological string theory on local Calabi--Yau threefolds. 

\end{titlepage}
\vfill
\eject

\tableofcontents

\sectiono{Introduction}

The perturbative partition function of a matrix model with fixed filling fractions $\epsilon$ has the form 
\be
\label{pertz}
Z=\exp \Bigl\{  \sum_{g=0}^{\infty} N^{2-2g} F_g(\epsilon)\Bigr\},
\eeq
where $F_g(\epsilon)$ is the generating function for fatgraphs of genus $g$ \cite{DJZ}. 
The same structure appears in the partition function of closed topological string theory, where $1/N$ becomes the string 
coupling constant $g_s$ and the $\epsilon$ become closed string moduli. In both cases, the partition function depends on a choice of background $\epsilon$.


The background dependence of $Z$ is closely related to its behavior under the modular group of the theory. 
In the context of matrix models, the large $N$ limit is described by an algebraic curve called the spectral curve, and the modular group is simply the symplectic group 
${\rm Sp}(2 \bar g,\IZ)$, where $\bar g$ is the genus of the spectral curve (not to be confused with the genera $g$ appearing in the topological 
expansion). In topological string theory on a Calabi--Yau threefold $X$, 
the modular group is the symplectic group ${\rm Sp}(2n, \IZ)$ of symmetries which preserve the 
symplectic form, where $n=b_3(X)/2$. 
The mathematical manifestation of background dependence is that, as emphasized in \cite{abk}, the 
partition function $Z$ does not have good transformation properties under the modular group of the theory. In fact, as shown in \cite{abk} in topological 
string theory and in \cite{eo} in the context of matrix models, the $F_g$ transform as quasi-modular forms, with shifts (the prototype for this behavior 
is the second Eisenstein series). 

It is possible to restore modularity of $F_g$ (hence of $Z$) at the price of introducing a non-holomorphic dependence on $\bar \epsilon$. 
When this is done, the resulting partition function $Z(\epsilon, \bar \epsilon)$ satisfies the holomorphic anomaly equations of \cite{bcov}. In the context 
of matrix models this was shown in \cite{emo}. Indeed, the holomorphic anomaly was interpreted in \cite{witten} as an obstruction to background independence. 
Therefore, the lack of background independence seems to face us with a choice between modularity and holomorphicity.

In the case of matrix models, it is however clear that the original 
matrix integral which leads to the above $1/N$ expansion can not depend on the choice of filling fractions. It should 
only depend on the 
coupling constants of the potential, the rank of the matrix $N$, and a choice of integration path for the eigenvalues. Therefore, for matrix models, 
background dependence is an artifact of the $1/N$ expansion, coming from the fact that one has chosen a particular saddle configuration at large $N$. Background 
independence should be restored by including the rest of the saddle points in an appropriate way, i.e. by including the instanton configurations of the matrix model. 
In this sense, matrix integrals provide a simple framework in which we might understand both the breakdown of background independence and the appropriate mechanism to 
restore it\footnote{See \cite{rozali} for an excellent review of background independence in field theory and string theory.}. 

In \cite{eone}, one of us proposed an asymptotic formula for the partition function of convergent matrix integrals, generalizing the 
results of \cite{bde}. This formula includes, together with the 
perturbative expansion (\ref{pertz}), a series of nonperturbative corrections which can be interpreted in terms of instantons of the matrix model. Since this nonperturbative partition function is obtained by summing over all possible filling fractions, it was suggested in \cite{eone} that it gives a natural proposal for a background independent partition function. 

As in the case of the $F_g$ \cite{eo}, the nonperturbative partition function introduced in \cite{eone} can be 
defined for {\it any} spectral curve $\Sigma$. 
On top of the spectral curve data, one also needs a choice of characteristics $(\mu,\nu)$, just as for theta functions on a Rieman surface.
These characteristics encode nonperturbative information; for example, in a matrix model they encode the choice of integration contour for the eigenvalues. In this paper we study in detail the transformation properties of the nonperturbative partition function $Z_{\Sigma} (\mu,\nu;\epsilon)$ under the modular group. It turns out that they have good modular properties. More precisely, they transform in a matrix representation 
of the modular group, i.e. 
\be
\label{tlaw}
\tilde Z_{\Sigma} (\tilde \mu,\tilde \nu;\tilde \epsilon)=\zeta\bigl[^\mu_\nu\bigr](\Gamma) Z_{\Sigma}( \mu,\nu;\epsilon),
\eeq
where $\zeta\bigl[^\mu_\nu\bigr](\Gamma)$ is a phase depending on the characteristics $\mu$, $\nu$ and the modular transformation $\Gamma$, and $\tilde \mu, \tilde \nu$ are new, transformed characteristics. Both $\zeta\bigl[^\mu_\nu\bigr](\Gamma)$ and $\tilde \mu, \tilde \nu$ are 
the same quantities which appear in the transformation properties of higher rank theta functions. Since $Z_{\Sigma}(\mu,\nu;\epsilon)$ is manifestly holomorphic, we obtain a partition function which is holomorphic, modular, and background independent. In other words, modularity can be restored in a holomorphic way by including nonperturbative effects.  
Notice that, according to (\ref{tlaw}), $Z_{\Sigma}(\mu,\nu;\epsilon)$ transforms like a twisted fermion on the Riemann surface, with twists given by the characteristics $\mu,\nu$. Therefore, it seems to be the most natural object from the point of view of the free fermion theory on $\Sigma$ advocated in \cite{adkmv,dhsv} and many other papers. As in CFT, one can also regard the $Z_{\Sigma}(\mu,\nu;\epsilon)$ as chiral conformal blocks which can be used to construct invariants under suitable subgroups of the modular group. 

We also show that the nonperturbative partition function is a tau function, in the sense that it satisfies a Hirota-type equation. This was observed in \cite{eo} in the case of genus zero spectral curve, 
where the nonperturbative instanton corrections are absent. Here we show that the inclusion of these corrections makes possible to generalize the construction of \cite{eo} 
to any spectral curve. 

Since the construction of $Z_{\Sigma} (\mu,\nu;\epsilon)$ only needs data coming from $\Sigma$, the nonperturbative partition function introduced in \cite{eone} 
can be also defined for topological strings on a variety of local Calabi--Yau manifolds, including mirrors of toric Calabi--Yau's. In fact, it has been advocated in \cite{msw,mmnp} that the nonperturbative topological string partition function should include matrix model-like instanton effects, and in \cite{mmnp} it was pointed out 
that the nonperturbative partition function of \cite{eone} appears naturally in topological string models with 
large $N$ Chern--Simons  theory duals. Therefore, these nonperturbative 
partition functions provide homolorphic, modular and background independent partition functions for 
a wide class of topological string theory models. 

The organization of this paper is as follows. In section 2 we recall the definition of the nonperturbative partition function introduced in \cite{eone}, and we stress the fact that it can be associated to any spectral curve. In section 3, which is the core of the paper, we show in detail that this partition function has good transformation properties under the modular group. In section 4 we review how the nonperturbative partition function appears in the context of matrix models, by summing over filling fractions. In section 5 we discuss the applications to topological string 
theory on local Calabi--Yau manifolds, and we propose that the nonperturbative partition function gives a natural nonperturbative object for topological strings. In section 6 we analyze the integrability properties and we show that $Z_{\Sigma} (\mu,\nu;\epsilon)$ is a tau function. Finally, in section 7 we list some conclusions and avenues for further research.

\sectiono{The nonperturbative partition function}


Consider an arbitrary ``spectral curve" $\Sigma=({\cal C},x,y)$, i.e. the data of a compact Riemann surface ${\cal C}$ of genus $\bar g$, together with two analytical functions $x,y$ on some open domain of ${\cal C}$. Its symplectic invariants $F_g$'s were defined in \cite{eo} (and we recall the definition in appendix \ref{appendixFg}). They are such that (if $g\geq 2$):
\be
F_g({\cal C},x,\l y) = \l^{2-2g}\,F_g({\cal C},x,y),
\eeq
 and if two spectral curves have the same symplectic form $\rd\td{x}\wedge \rd\td{y}=\rd x\wedge \rd y$, we have 
 \be
 F_g({\cal C},\td{x}, \td{y})=F_g({\cal C},x, y).
 \eeq

\smallskip
For the spectral curve $\Sigma=({\cal C},x,y)$, the {\it nonperturbative partition function} introduced in \cite{eone} in the context of matrix models, by summing over filling fractions (see section \ref{sectionMM} for more details on the origin of this definition), is defined by
\be
\label{exphatZTheta}
\ba
& Z_\Sigma (\mu,\nu;\epsilon)
= \ee{\sum_{g\geq 0} N^{2-2g} F_g(\epsilon)}  \sum_{k} \sum_{l_i>0}\sum_{h_i>1-{l_i\over 2}} {N^{\sum_i (2-2h_i-l_i)}\over k! l_1!\,\dots\, l_k!}\,\,\, F_{h_1}^{(l_1)}\dots F_{h_k}^{(l_k)}  \,\, \Theta_{\mu,\nu}^{(\sum_i l_i)}(NF'_0,\tau) \\
&\quad = \ee{N^2 F_0}\, \ee{F_1}\, \ee{(N^{-2} F_2 + N^{-4} F_3 +\dots)} \,\, \biggl\{ \Theta_{\mu,\nu} + {1\over N}\Bigl(\Theta'_{\mu,\nu} F_1' + {1\over 6} \Theta_{\mu,\nu}'''\,F_0'''\Bigr)  \\
& \quad +{1\over N^2}\Bigl({1\over 2}\Theta_{\mu,\nu}'' F_1'' + {1\over 2}\Theta_{\mu,\nu}'' F_1'^2 + {1\over 24}\Theta_{\mu,\nu}^{(4)} F_0''''+{1\over 6}\Theta_{\mu,\nu}^{(4)} F_0''' F_1' + {1\over 72}\Theta_{\nu,\mu}^{(6)} F_0'''^2 \Bigr) + \dots \biggr\}. 
\ea
\eeq
In this partition function, the $F_g$'s are the symplectic invariants \cite{eo} of the spectral curve $\Sigma$, their derivatives are with respect to the background filling fraction $\epsilon$ and computed at:
\beq
\label{eps}
\epsilon = {1\over 2\pi \ri}\,\oint_{\acycle} y  \rd x,
\eeq
and the $F_g({\cal C},x,y)$'s and their derivatives depend on a choice of symplectic basis of $2\bar{g}$ one-cycles ${\cal A}_i,{\cal B}_j$ on ${\cal C}$. Finally, the theta function $\Theta_{\mu,\nu}$ of characteristics $(\mu,\nu)$ is defined by
\be
\label{biget}
\Theta_{\mu,\nu}(u,\tau) = \sum_{n\in {\mathbb Z}^{\bar g}} \ee{(n+\mu-N \epsilon)u}\,\,\ee{\pi \ri  (n +\mu-N\epsilon)\tau (n+\mu-N\epsilon)}\,\,\ee{2 \ri\pi n \nu}
\eeq
and is evaluated at
\be\label{defutau}
u=N F_0',
\qquad
F'_0  = \oint_{\bcycle} y(x) \rd x,
\qquad
\tau =  {1\over 2\pi \ri} F_0''.
\eeq
In (\ref{exphatZTheta}), the derivatives of the theta function (\ref{biget}) are w.r.t. $u$, therefore 
each derivative introduces a factor of $n+\mu-N\epsilon$ in the sum (\ref{biget}).
The derivatives of $\Theta$ and the derivatives of $F_g$, are written with tensorial notations. 
For instance, ${1\over 6}\Theta_{\mu,\nu}^{(4)} F_0''' F_1' $ actually means:
\beq
{1\over 6}\Theta_{\mu,\nu}^{(4)} F_0''' F_1' 
\equiv {1\over 2!\, 3!\, 1!}\,\, \sum_{i_1,i_2,i_3,i_4}\, {\partial^4 \Theta_{\mu,\nu}\over \partial u_{i_1}\partial u_{i_2}\partial u_{i_3}\partial u_{i_4}}\,\, {\partial^3 F_0\over \partial \epsilon_{i_1}\partial \epsilon_{i_2}\partial \epsilon_{i_3}}\,\, {\partial F_1\over \partial \epsilon_{i_4}}
\eeq
and the symmetry factor (here ${1\over 6}={2\over 2!\,3!\,1!}$) is the number of relabellings of the indices, giving the same pairings, and divided by the order of the group of relabellings, i.e. $k!\, l_1!\dots l_k!$, as usual in Feynmann graphs.




The $\Theta$ function is closely related to the standard theta function, which is defined by 
\be
\vartheta\bigl[^\mu_\nu\bigr](\xi|\tau)= \sum_{{\bf n}\in {\mathbb Z}^{\bar g}}
\exp\bigl[ \ri \pi (n+\mu) \tau (n+\mu) + 2\pi \ri (n+\mu)(\xi+\nu)\bigr].
\eeq
It it easy to see that these two functions are related as follows
\be
\label{Thetatheta}
\Theta_{\mu,\nu}(u,\tau) =\exp\biggl[ -N^2\Bigl(  \epsilon F_0'+{1\over 2} \epsilon^2 F_0'' \Bigr)\biggr] 
\,\, \vartheta\bigl[^\mu_\nu\bigr](\xi|\tau)
\eeq
where
\be
\label{xivalue}
\xi={N \over 2\pi \ri}  \oint_{\bcycle-\tau\acycle} y (x) \rd x =N \biggl( {F'_0 \over 2\pi \ri}  - \tau \epsilon \biggr).
\eeq

Finally, we point out that, as shown in \cite{eone}, the $1/N$ corrections in (\ref{exphatZTheta}) can be resummed in terms of a single theta function, and $Z_{\Sigma}(\mu,\nu;\epsilon)$ can be written as
\be
\label{resumZ}
Z_{\Sigma}(\mu,\nu;\epsilon)=\ee{\sum_{g\geq 0} N^{2-2g} F_g(\epsilon)} \Theta_{\mu, \nu}(u^N, \tau^N), 
\eeq
where 
\be
u^N=NF_0'\bigl(1+\CO(1/N^2)\Bigr), \quad \tau^N=\tau +\CO(1/N^2),
\eeq
and the corrections (which depend on the $F_g$) can be easily determined order by order in $1/N$. 

\sectiono{Modular properties of the nonperturbative partition function}

In this section we discuss the transformation properties of the nonperturbative partition function (\ref{exphatZTheta}) 
under the modular group. For that purpose, we shall first remind the transformation properties of the $F_g$'s and their derivatives, and then the transformation properties of the $\Theta$-function. For simplicity, we will first discuss the case in which $\bar g=1$ and the 
theta functions involved are of rank one. There are little changes when we go to the general case, but we will indicate these 
in section 3.3 below.

\medskip

For spectral curves of $\bar g=1$, a general modular transformation is an element of $ {\rm SL}(2,\IZ)$,
\be
\Gamma=\begin{pmatrix}\alpha&\beta\\ \gamma & \delta \end{pmatrix} \in {\rm SL}(2,\IZ), 
\eeq
where $\alpha, \beta, \gamma, \delta \in \IZ$ and $\alpha\delta-\gamma\beta=1$. 
Under this transformation the integrals of $y \rd x$ over the $\CB$ and the $\CA$ cycles transform as
\beq
\begin{pmatrix}F'_0 \\ 2\pi \ri \epsilon \end{pmatrix} \rightarrow \begin{pmatrix}\td{F}'_0 \\ 2\pi \ri \td\epsilon\end{pmatrix} = \begin{pmatrix}\alpha&\beta\\ \gamma & \delta \end{pmatrix}\,\begin{pmatrix}F'_0 \\ 2\pi \ri \epsilon \end{pmatrix}.
\eeq
It follows that
\beq
\td\tau ={ \alpha\tau+\beta\over \gamma\tau+\delta}, \qquad \tilde \xi = {\xi \over \gamma\tau+\delta},
\eeq
so in particular $\xi$ is a modular form of weight $-1$.

\subsection{Modular transformations of the perturbative amplitudes}

The transformation properties of the genus zero free energy $F_0$ can be derived from those of $F_0'$ by integration, and they read
\be
\tilde F_0= F_0+{1\over 2} \delta \beta (2\pi \ri \epsilon)^2 +{1\over 2} \gamma \alpha (F_0')^2 + \beta \gamma (2\pi \ri \epsilon) F_0'.
\eeq
This implies that the combination 
\beq
\label{invariant}
\td{F}_0 - {1\over 2}\td\epsilon\,\td{F}'_0 = F_0 - {1\over 2}\epsilon\, F'_0
\eeq
stays invariant\footnote{The invariance of (\ref{invariant}) is a well-known fact in the context of Seiberg--Witten theory \cite{matone}, where this combination turns out to be the modulus of the Seiberg--Witten curve. This invariance can also be seen from the explicit expression of $F_0$ in \cite{eo}.}. 

Another transformation property which will be useful in the following is that
\be
\label{shifted}
2 \pi \ri \epsilon (F_0'- 2\pi \ri \epsilon \tau) \rightarrow  2 \pi \ri \epsilon (F_0'- 2\pi \ri \epsilon \tau) -\kappa 
(F_0'- 2\pi \ri \tau \epsilon )^2,
\eeq
where
\beq
\kappa = -(\gamma\tau+\delta)^{-1}\, \gamma.
\eeq
\begin{figure}[!ht]
\leavevmode
\begin{center}
\epsfxsize=4.5cm
\epsfysize=6cm
\epsfbox{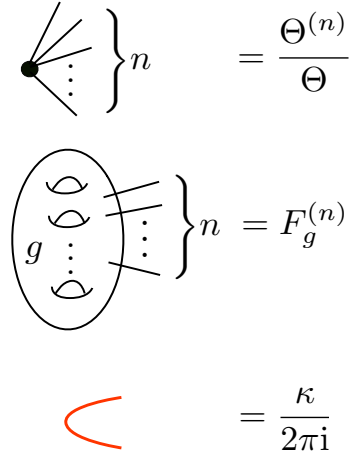}
\end{center}
\caption{We will represent the different ingredients appearing in the modular transformation of the nonperturbative partition function by the graphic symbols depicted above.}
\label{fig:modulardic}
\end{figure}
The transformation properties for the $F_g$, $g\ge 1$ were studied in \cite{abk} from the point 
of view of topological string theory. \cite{abk} also found an elegant diagrammatic formalism to express these 
transformations. The formalism can be depicted as follows:
after a modular transformation, $F_g$ gets contributions corresponding to all possible stable degenerations of a genus $g$ Riemann surface. Degenerations are obtained by pinching a non-trivial cycle. A stable degenerate surface is a nodal surface with marked points whose components have strictly negative Euler characteristics.
Each degenerate cycle becomes a nodal point, and carries a factor $(2\ri\pi)^{-1}\kappa$. For example, the modular transformation 
of $F_2$ gets contributions corresponding to pinching either 1,2 or 3 cycles. This leads to the following transformation properties for the $F_g$:
\be
\label{tfg}
\ba
\td{F}_1 &= F_1 -{1\over 2} \ln (\gamma \tau+\delta),\\
\td{F}_2 &= F_2 +  {\kappa \over 2\pi \ri} {1\over 2} (F_1''+{F_1'}^2)+{1\over 8}\Bigl({\kappa\over 2\pi \ri}\Bigr)^2 (F_0''''+4F_0'''F_1')+{1\over 48}\Bigl({\kappa\over 2\pi \ri}\Bigr)^3(6 {F'_0}^3 + 4 {F'_0}^3)
\ea
\eeq
and so on. For $F_2$, in the higher rank $\bar g>1$ case, the last two terms have a different index structure, as it can be seen from the graphical 
representation, and this is why we have written them in separate form, anticipating our general analysis. In general for $g\geq 2$, $\td{F}_g$ is a polynomial 
in $\kappa$ of degree $3g-3$. This method can be extended (see for example \cite{eo}) to include all derivatives of the $F_g$, 
and the derivatives correspond simply to insertions of marked points. One should also note that the transformation of $F_g^{(n)}$ includes an overall factor of 
\be
(\gamma \tau + \delta)^{-n}.
\eeq

\begin{figure}[!ht]
\leavevmode
\begin{center}
\epsfxsize=14cm
\epsfysize=3cm
\epsfbox{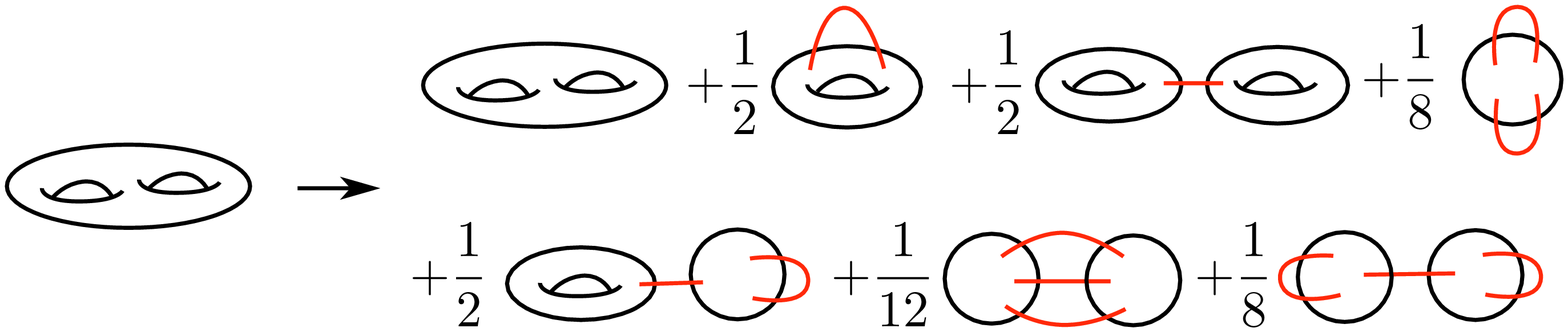}
\end{center}
\caption{A graphical depiction of the modular transformation of $F_2$.}
\label{f2modular}
\end{figure}

\begin{figure}[!ht]
\leavevmode
\begin{center}
\epsfxsize=10cm
\epsfysize=1.5cm
\epsfbox{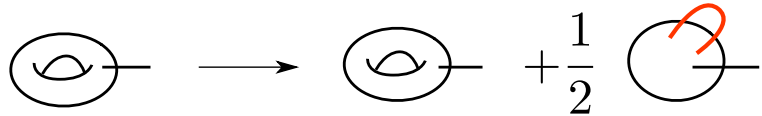}
\end{center}
\caption{A graphical depiction of the modular transformation of $F_1'$.}
\label{genusoneder}
\end{figure}

In the following, we will rely very heavily on diagrammatic methods, and we will use the graphical representation shown in \figref{fig:modulardic} for the different ingredients appearing in the calculations. In this diagrammatic language, the modular transformation of $F_2$ can be 
represented as in \figref{f2modular}, while in \figref{genusoneder} we show the transformation 
properties of $F_1'$. 
The numerical factors in front of each diagram are the symmetry factors of the corresponding diagram, i.e. the number of ways of pinching giving the same diagram, and divided by the order of the automorphism goup. 
This is the usual symmetry factor of Feynman graphs.

\begin{figure}[!ht]
\leavevmode
\begin{center}
\epsfxsize=8cm
\epsfysize=10cm
\epsfbox{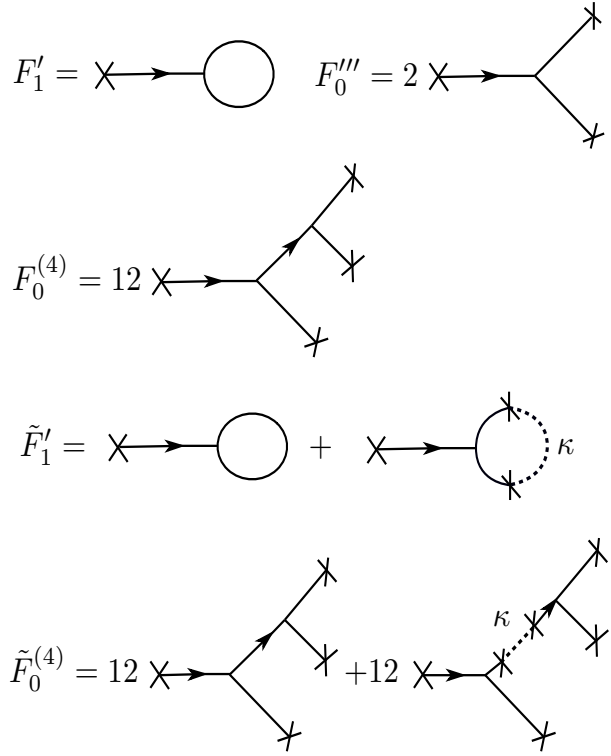}
\end{center}
\caption{A graphical depiction of the modular transformation of $F_1'$  and $F_0^{(4)}$ using the diagrammatic 
representation of \cite{eynloopeq,eo}. Each arrowed edge means a propagator $K$ in \cite{eo}, and each non-arrowed edge means a Bergmann kernel. Each cross at the end of edges, means that we take the ${\cal B}_i$ cycle integral corresponding to $\partial/\partial \epsilon_i$. For example $\partial F_1/\partial\epsilon_i = \oint_{z\in{\cal B}_i}\, \sum_j \Res_{z'\to a_j}\, K(z,z')B(z',\bar{z}') $. Modular transformations amount to cutting edges in all possible ways such that each subdiagram contains at least one vertex.}
\label{eomodular}
\end{figure}

In the context of matrix models, the  
transformation properties of the $F_g$ can be derived from the formalism of \cite{eo}. 
The basic ingredient in this formalism  is the transformation property of the 
Bergmann kernel under a modular transformation, 
\beq
\td{B}(z_1,z_2)  = B(z_1,z_2) + 2\ri \pi \om(z_1)\,\kappa\,\om(z_2),
\eeq
where $\omega(z)$ is a basis of Abelian holomorphic differentials on $\Sigma$.
The $F_g$'s of \cite{eo} are made of residues of products of Bergmann kernels (see appendix \ref{appendixFg}), namely $F_g^{(n)}$ is a sum of residues of products of $3g-3+n$ Bergmann kernels, and multiplied by terms independent of a choice of cycles.
Each term of $F_g^{(n)}$ is thus represented in \cite{eo} by a trivalent diagram, with $3g-3+n$ edges.
A modular transformation amounts to cutting (or nor cutting) edges of each 
diagram in all possible ways (such that all subdiagrams have at least one vertex), and 
each cut edge is replaced by a factor of $\kappa$. We illustrate these rules in \figref{eomodular}. One has for example 
\be
\td{F}_0^{(4)} = F_0^{(4)} + 3{\kappa\over 2\ri\pi} (F_0''')^2,
\eeq
Thus, the modular transformation of $F_g^{(n)}$ can be written as a polynomial of degree $3g-3+n$ of $\kappa$.
One can show \cite{emo}, using this property, that one obtains the same diagrammatic calculus as in \cite{abk}.

\subsection{Transformation properties of the theta function}

We now study the transformation properties of the $\Theta$-function, and first, we remind the transformation properties of the standard theta function, 
\be
\label{thetatrans}
\tilde \vartheta\bigl[^{\tilde \mu}_{\tilde \nu}\bigr] (\tilde \xi |\tilde \tau)= \zeta\bigl[^\mu_\nu\bigr](\Gamma) (\gamma \tau+\delta)^{{1\over 2}} \exp\Bigl[ -\pi \ri \kappa \, \xi^2\Bigr]\vartheta\bigl[^\mu_\nu \bigr](\xi|\tau), 
\eeq
where
\be
\ba
\tilde \mu &=\delta \mu - \gamma \nu +{1\over 2} \gamma \delta,\\
\tilde \nu &=-\beta \mu + \alpha \nu +{1\over 2}\alpha \beta,
\ea
\eeq
and
\be
 \zeta\bigl[^\mu_\nu\bigr](\Gamma)=\exp \biggl\{-\pi \ri  \Bigl(\delta \beta \mu^2 + \gamma \alpha \nu^2 -2 \beta \gamma \mu \nu 
 +\alpha \beta (\delta \mu -\gamma \nu) \Bigr) \biggr\}  \zeta\bigl[^0_0\bigr](\Gamma).
\eeq
Here, $\zeta\bigl[^0_0\bigr](\Gamma)$ is a root of unity which does not depend on the characteristics.

 We now deduce an important transformation property of the derivatives of the theta function 
which will be useful in the following. The derivatives appearing in (\ref{exphatZTheta}) can be 
computed in terms of the standard theta function 
by considering
\be
{1\over (2 \pi \ri)^{\ell}}\Bigl( \partial_{\xi}- N 2 \pi \ri \epsilon \Bigr)^{\ell} \vartheta\bigl[^\mu_\nu\bigr](\xi|\tau), 
\eeq
and we are interested on the transformation properties of such quantities under a modular 
transformation. In evaluating these quantities, the variable $\xi$ is regarded as an independent 
variable, unrelated to $\epsilon$ and transforming as a modular form of weight $-1$. Only at the very end we set it equal to its true value. Under 
a modular transformation, 
\be
\label{derthet}
\Bigl( \partial_{\tilde \xi}- N 2 \pi \ri \tilde \epsilon \Bigr)^{\ell} \tilde \vartheta\bigl[^{\tilde \mu}_{\tilde \nu}\bigr](\tilde \xi| \tilde \tau) =
\CN  \Bigl(\partial_{\tilde \xi}- N 2 \pi \ri \tilde \epsilon \Bigr)^{\ell} \exp\Bigl[ -\pi \ri \kappa \xi^2 \Bigr] \vartheta\bigl[^{\mu}_{\nu}\bigr](\xi|\tau)
\eeq
where $\CN$ does not depend on $\xi$. We now evaluate 
\be
A_{\ell}(f)=\exp\Bigl[ \pi \ri \kappa \xi^2 \Bigr]  \Bigl( \partial_{\tilde \xi}- N 2 \pi \ri \tilde \epsilon \Bigr)^{\ell}\exp\Bigl[ -\pi \ri \kappa \xi^2 \Bigr] f(\xi)
\eeq
where $f(\xi)$ is an arbitrary function of $\xi$. We find, 
\be
A_{\ell}(f)=\Bigl( \partial_{\tilde \xi}+2\pi \ri \gamma \xi- N 2 \pi \ri \tilde \epsilon \Bigr)^{\ell}  f(\xi)
\eeq
The operators appearing in the r.h.s. do not commute, and it is easier to consider a generating 
functional
\be
S(x) =\sum_{\ell=0}^{\infty} {A_{\ell}(f) \over \ell!} x^{\ell}=\re^{x(\partial_{\tilde \xi}+2\pi \ri \gamma \xi- N 2 \pi \ri \tilde \epsilon) }  f(\xi).
\eeq
Using the Baker--Campbell--Hausdorff formula, we get
\be
S(x)= \re^{x(2\pi \ri \gamma \xi- N 2 \pi \ri \tilde \epsilon) } 
\re^{-x^2 \pi \ri \kappa} \re^{x \partial_{\tilde \xi}} f(\xi).
\eeq
After extracting the $\ell$-th power of $x$ in this generating functional, we can already set $\xi$ to its value (\ref{xivalue}). Since 
\be
2\pi \ri \gamma \xi - N 2 \pi \ri \tilde \epsilon =-(\gamma \tau +\delta) N 2\pi \ri  \epsilon
\eeq
we finally obtain
\be
A_{\ell}(f)=(c\tau +\delta)^{\ell} \sum_{j=0}^{[\ell/2]} (2j-1)!! {\ell \choose 2j} (-2\pi \ri \kappa)^j \bigl(\partial_\xi-N 2 \pi \ri  \epsilon \bigl)^{\ell-2j} f(\xi),
\eeq
and this implies the following transformation law for the derivatives of the theta function,
\be
\label{thetader}
\ba
&{1\over \tilde \vartheta\bigl[^{\tilde \mu}_{\tilde \nu}\bigr](\tilde \xi|\tilde \tau)} \Bigl( {1\over 2\pi \ri} \partial_{\tilde \xi}- N \epsilon \Bigr)^{\ell} \tilde \vartheta\bigl[^{\tilde \mu}_{\tilde \nu}\bigr](\tilde \xi|\tilde \tau) =\\
& {1\over \vartheta\bigl[^{\mu}_{\nu}\bigr](\xi|\tau)} 
(\gamma \tau+\delta)^{\ell} \sum_{j=0}^{[\ell/2]} (2j-1)!! {\ell \choose 2j} \Bigl( -{\kappa \over 2\pi \ri} \Bigr)^j \Bigl( {1\over 2\pi \ri}  \partial_\xi- N \epsilon\Bigr)^{\ell-2j} \vartheta\bigl[^{ \mu}_{ \nu}\bigr](\xi|\tau).
\ea
\eeq

\subsection{Generalization to higher rank}

In the higher rank case, the modular group is ${\rm Sp}(2\bar g, \IZ)$. A modular transformation $\Gamma$ satisfies
\be
\Gamma^{\rm T} \Omega 
\Gamma=\Omega, \qquad \Omega=\begin{pmatrix}0& {\bf 1} \\
               -{\bf 1}&0\end{pmatrix},
               \eeq
and it can be written as:
\be
\Gamma=\begin{pmatrix}A& B \\
               C&D \end{pmatrix} 
\label{simp}
\eeq
where the $\bar g \times \bar g$ matrices $A$, $B$, $C$, $D$, with integer-valued entries, satisfy
\be
A^{\rm T}D-C^{\rm T}B= {\bf 1}_{\bar g}, \,\,\,\ 
A^{\rm T}C=C^{\rm T}A, \,\,\,\ 
B^{\rm T}D=D^{\rm T}B.
\label{simple}
\eeq
All previous quantities are promoted to vectors and matrices, with obvious generalizations. For example, 
\be
\ba
\td{F}'_{0,j}&=A_j^{~k} F'_{0,k}+B_{jk} 2\pi \ri \epsilon^k,\\  
2\pi \ri \td\epsilon^j&=C^{jk} F'_{0,k}+ D^j_{~k} 2\pi \ri \epsilon^k,
\ea
\eeq
where summation over repeated indices is understood, and
\beq
\td\tau = (A\tau+B)(C\tau + D)^{-1}, \qquad \tilde \xi_i = \Bigl[ (C\tau+D)^{-1}\Bigr]^j_{~i} \xi_j, \quad i=1, \cdots, \bar g. 
\eeq
The genus zero free energy  transforms now as
\be
\tilde F_0=F_0 +
{1 \over 2} 2\pi \ri \epsilon ^k(D^{\rm T} B)_{kj}2\pi \ri \epsilon^j+ {1 \over 2} F'_{0,k} 
(C^{\rm T}A)^{kj}F'_{0,j}+ 2\pi \ri \epsilon^k (B^{\rm T}C)_k^{~j} F'_{0,j}.
\label{premono}
\eeq
The quantity
\be
\label{geninv}
F_0 - {1\over 2}\epsilon\, F'_0 \equiv F_0- {1\over 2} \epsilon^k \, F'_{0,k}
\eeq
is still invariant, as it can be easily checked. In the following, for general $\bar g$, the expressions we used for $\bar g=1$ will 
denote the obvious contraction of 
indices, as in (\ref{geninv}) above. We also have the obvious generalization of (\ref{shifted}),
\be
\label{genshifted}
2 \pi \ri \epsilon (F_0'- 2\pi \ri \epsilon \tau) \rightarrow  2 \pi \ri \epsilon (F_0'- 2\pi \ri \epsilon \tau) -(F_0'- 2\pi \ri \tau \epsilon )\kappa 
(F_0'- 2\pi \ri \tau \epsilon ), 
\eeq
where 
\be
\kappa^{ij}= -\Bigl[ (C\tau + D)^{-1} C\Bigr]^{ij}. 
\eeq
This matrix is {\it symmetric}, as a consequence of the symplectic properties of $\Gamma$. 

The transformation properties of the $F_g$ are given by the diagrammatic method explained above. We note that, for $\bar g>1$, 
\be
\label{genone}
\tilde F_1= F_1 -{1\over 2} \log \, {\rm det}(C\tau +D).
\eeq

Let us now consider the transformation properties of the theta function and its derivatives, in the higher genus case. We have (see 
for example \cite{agmv})
\be
\label{thetatransgen}
\tilde \vartheta\bigl[^{\tilde \mu}_{\tilde \nu}\bigr] (\tilde \xi |\tilde \tau)= \zeta\bigl[^\mu_\nu\bigr](\Gamma)\Bigl( 
{\rm det}(C \tau+D)\Bigr) ^{{1\over 2}} \exp\Bigl[ -\pi \ri \xi \kappa \xi \Bigr]\vartheta\bigl[^\mu_\nu \bigr](\xi|\tau).
\eeq
In this transformation, the characteristics are given by 
\be
\label{charchange}
\ba
\tilde \mu &=D \mu -C \nu +{1\over 2} (CD^T)_{\rm d},\\
\tilde \nu &=-B \mu + A \nu +{1\over 2}(AB^T)_{\rm d}.
\ea
\eeq
where $(M)_{\rm d}$ denotes a column vector whose entries are the diagonal entries of the matrix $M$. The phase is now given by 
\be
 \zeta\bigl[^\mu_\nu\bigr](\Gamma)=\exp \biggl\{-\pi \ri  \Bigl(\mu D^T B \mu +\nu C^T A \nu -2 \mu B^T C \nu 
 +(\mu D^T -\nu C^T) (AB^T)_{\rm d} \Bigr) \biggr\}  \zeta\bigl[^0_0\bigr](\Gamma).
\eeq

We now consider the derivatives of the theta function. We want to generalize (\ref{thetader}) and to compute the modular transformation of 
\be
\Bigl( \partial^{j_1}- N 2 \pi \ri \epsilon^{j_1} \Bigr) \cdots \Bigl( \partial^{j_\ell}- N 2 \pi \ri \epsilon^{j_\ell} \Bigr) \vartheta\bigl[^\mu_\nu\bigr](\xi|\tau), 
\eeq
where
\be
\partial^j ={\partial \over \partial \xi_j}.
\eeq
Under 
a modular transformation, 
\be
\label{derthetgen}
\ba
&\Bigl( \tilde \partial^{j_1} - N 2 \pi \ri \tilde \epsilon^{j_1} \Bigr)   \cdots \Bigl(\tilde \partial^{j_\ell}- N 2 \pi \ri \tilde \epsilon^{j_\ell} \Bigr) \vartheta\bigl[^{\tilde \mu}_{\tilde \nu}\bigr](\tilde \xi| \tilde \tau) =\\
& \CN   \Bigl( \tilde \partial^{j_1} - N 2 \pi \ri \tilde \epsilon^{j_1} \Bigr)  \cdots \Bigl(\tilde \partial^{j_\ell}- N 2 \pi \ri \tilde \epsilon^{j_\ell} \Bigr)\exp\Bigl[ -\pi \ri \xi \kappa \xi \Bigr] \vartheta\bigl[^{\mu}_{\nu}\bigr](\xi|\tau)
\ea
\eeq
where $\CN$ is again independent of $\xi$. Since 
\be
\Bigl( \tilde \partial^{j} - N 2 \pi \ri \tilde \epsilon^{j} \Bigr)\exp\Bigl[ -\pi \ri \xi \kappa \xi \Bigr]= \exp\Bigl[ -\pi \ri \xi \kappa \xi \Bigr] 
\Bigl( \tilde \partial^{j} +2\pi \ri C^{jk}\xi_k- N 2 \pi \ri \tilde \epsilon^{j} \Bigr)
\eeq
we just have to compute
\be
\label{optheta}
A^{j_1} \cdots A^{j_\ell} \, \vartheta\bigl[^{\mu}_{\nu}\bigr](\xi|\tau). 
\eeq
where the operator $A^j$ is given by
\be
A^j =(C\tau + D)^j_{~k} \partial^{k} +2\pi \ri C^{jk}\xi_{k}- N 2 \pi \ri \tilde \epsilon^{j}.
\eeq
As it happened in the $\bar g=1$ case, the operators $A^{j_1}$, $\cdots$, $A^{j_{\ell}}$ do not commute, since we cannot set $\xi_k$ to its value until we have not commuted all the derivatives to the right. Of course, this is precisely the type of computation that Wick's theorem does. In this context, we define a normal-ordered operator as an operator in which we set 
\be
\xi_k=N \Bigl( {F_{0,k}' \over 2\pi \ri}- \tau_{kl}\epsilon^l \Bigr), 
\eeq
i.e. we have
\be
: A^{j_1} \cdots A^{j_\ell}: \, = (C\tau + D)^{j_1}_{~k_1} \cdots (C\tau + D)^{j_\ell}_{~k_\ell} (\partial^{k_1} - N 2\pi \ri \epsilon^{k_1}) \cdots
(\partial^{k_\ell} - N 2\pi \ri \epsilon^{k_\ell}).
\eeq
The contraction is given by
\be
\langle A^{j} A^k\rangle= 2\pi \ri C^{jk}= (C\tau + D)^{j}_{~l}(-2\pi \ri \kappa)^{lk}. 
\eeq
We can now apply Wick's theorem to write the operator in (\ref{optheta}) as a sum of normal-ordered operators. In the end we obtain
\be
\label{thetadergen}
\ba
&{1\over \tilde \vartheta\bigl[^{\tilde \mu}_{\tilde \nu}\bigr](\tilde \xi| \tilde \tau)}\Bigl( {1\over 2\pi \ri} \tilde \partial^{p_1} - N  \tilde \epsilon^{p_1} \Bigr)   \cdots \Bigl({1\over 2\pi \ri} \tilde \partial^{p_\ell}- N  \tilde \epsilon^{p_\ell} \Bigr) \vartheta\bigl[^{\tilde \mu}_{\tilde \nu}\bigr](\tilde \xi| \tilde \tau) =\\
&{1\over \vartheta\bigl[^{ \mu}_{ \nu}\bigr](\xi|\tau)}  (C\tau + D)^{p_1}_{~q_1} \cdots (C\tau + D)^{p_\ell}_{~q_\ell} 
\sum_{j=0}^{[\ell/2]} \sum_{\sigma} \Bigl(-{ \kappa \over 2\pi \ri} \Bigr)^{q_{\sigma(1)}q_{\sigma(2)} } \cdots \Bigl( -{ \kappa \over 2\pi \ri} \Bigr)^{q_{\sigma(2j-1)} q_{\sigma(2j)}} \\ 
& \Bigl( {1\over 2\pi \ri} \partial^{q_{\sigma(2j+1)}} - N \epsilon^{q_{\sigma(2j+1)}} \Bigr) \cdots
\Bigl( {1\over 2\pi \ri}  \partial^{q_{\sigma(\ell)}} - N \epsilon^{q_{\sigma(\ell)}}\Bigr)  \vartheta\bigl[^{ \mu}_{ \nu}\bigr](\xi|\tau).
\ea
\eeq
In the r.h.s., the first sum is over the number $j$ of contractions. The second sum is over all possible ways of performing the contractions, and 
$\sigma$ denotes the permutation of the indices $1, \cdots, \ell$ associated to a given contraction (note that not all possible permutations appear). (\ref{thetadergen}) is the 
generalization of (\ref{thetader}) to the multi-index case. Indeed, the combinatorial factor in the r.h.s. of (\ref{thetader}) is simply 
the number of possible ways of performing $j$ contractions in $\ell$ terms: the combinatorial number accounts for the choices of $2j$ legs to be contracted 
among $\ell$ legs in total, and $(2j-1)!!$ is the number of possible pairings of the $2j$ legs. 

\begin{figure}[!ht]
\leavevmode
\begin{center}
\epsfxsize=10cm
\epsfysize=7cm
\epsfbox{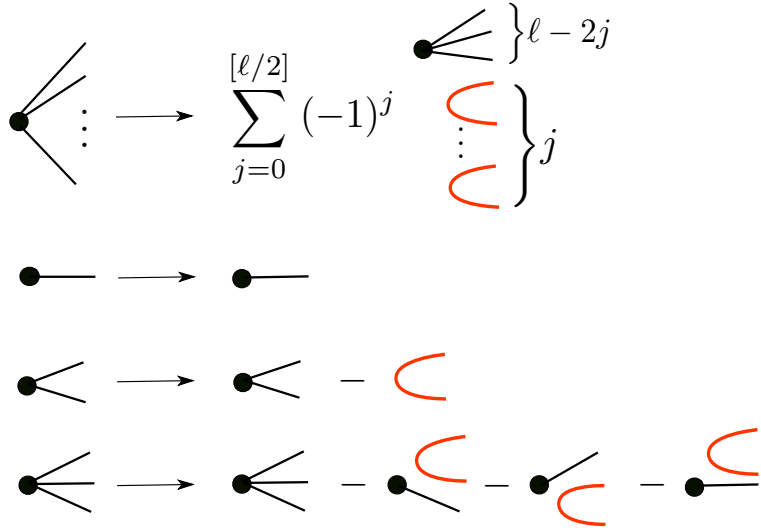}
\end{center}
\caption{The first line gives a graphical representation of (\ref{thetadergen}), while the next lines exemplify it for $\ell=1,2,3$. Notices that the 
edges are labeled, and when summing over all possible contractions we have to take this labeling into account, as shown in the last example.}
\label{fig:thetader}
\end{figure}

Of course, the easiest way of keeping track of combinatorial formulae like the above is by means of a graphical representation. Using the 
building blocks shown in \figref{fig:modulardic}, 
the equation (\ref{thetadergen}) might be represented as in \figref{fig:thetader}, where some simple examples are also shown. 

\subsection{Transformation properties of the nonperturbative partition function}

We will now study in detail the transformation properties of the nonperturbative partition function (\ref{exphatZTheta}). We rewrite 
(\ref{exphatZTheta}) as:
\beq\label{exphatZTheta2}
\ba
Z_{\Sigma}(\mu,\nu;\epsilon)
&= \ee{ N^{2} F_0+F_1} \, \Theta_{\mu,\nu} \, 
 \biggl\{ 1+ \sum_{j=1}^\infty N^{-j}\,Z_j(\mu,\nu;\epsilon)\biggr\} \\
%
%
&= \ee{ N^{2} F_0 +F_1} \, \Theta_{\mu,\nu} \, 
 \biggl\{ 1+ 
{1\over N}\Bigl({\Theta'_{\mu,\nu}\over \Theta_{\mu,\nu}} F_1' + {1\over 6} {\Theta_{\mu,\nu}'''\over \Theta_{\mu,\nu}}\,F_0'''\Bigr)  \\
& +{1\over N^2}\Bigl(F_2 + {1\over 2}{\Theta''_{\mu,\nu}\over \Theta_{\mu,\nu}} F_1'' + {1\over 2}{\Theta''_{\mu,\nu}\over \Theta_{\mu,\nu}} F_1'^2
+ {1\over 24}{\Theta^{(4)}_{\mu,\nu}\over \Theta_{\mu,\nu}} F_0'''' 
+ {1\over 6}{\Theta^{(4)}_{\mu,\nu}\over \Theta_{\mu,\nu}} F_0''' F'_1
+ {1\over 72}{\Theta^{(6)}_{\mu,\nu}\over \Theta_{\mu,\nu}} (F_0''')^2\Bigr )\\
& + \dots
\biggr\} \\
\ea
\eeq
i.e.  $Z_j$ is the sum of all terms contributing to order $N^{-j}$. 

We first study the transformation properties of the leading term $\ee{ N^{2} F_0 +F_1} \, \Theta_{\mu,\nu}$. To do this, we use the 
relation (\ref{Thetatheta}) to write
\be
\re^{N^2 F_0 } \Theta_{\mu,\nu}=\exp\biggl[ N^2\Bigl( F_0- \epsilon F_0'-{1\over 2} \epsilon^2 F_0'' \Bigr)\biggr] \vartheta\bigl[^\mu_\nu\bigr](\xi|\tau).
\eeq
In order to study the transformation properties of this quantity, we have to study the transformation 
properties of 
\be
\Xi=  F_0- \epsilon F_0'-{1\over 2} \epsilon^2 F_0'' =  F_0- {1\over 2} \epsilon F_0' - {1\over 2} \epsilon 
\Bigl( F'_0- 2\pi \ri \epsilon \tau \Bigr). 
\eeq
Using (\ref{genshifted}) and the invariance of (\ref{geninv}), one finds that
\be
\widetilde \Xi = \Xi + \pi \ri  \biggl( {F'_0 \over 2\pi \ri}  - \tau \epsilon \biggr) \kappa \biggl( {F'_0 \over 2\pi \ri}  - \tau \epsilon \biggr), 
\eeq
therefore the shift in $\Xi$ exactly compensates the $\xi$ dependent exponent in the transformation 
(\ref{thetatransgen}) and we find, 
\be
\re^{N^2 \tilde F_0 } \widetilde \Theta_{\tilde \mu,\tilde \nu} (\tilde u,\tilde \tau)=  \zeta\bigl[^\mu_\nu\bigr](\Gamma) \bigl({\rm det}(C\tau +D)\bigr)^{{1\over 2}}\re^{N^2 F_0 } \Theta_{ \mu, \nu} (u,\tau).
\eeq
It follows immediately from (\ref{genone}) that
\be
\re^{N^2 \tilde F_0 + \tilde F_1} \widetilde \Theta_{\tilde \mu,\tilde \nu} (\tilde u,\tilde \tau) =\zeta\bigl[^\mu_\nu\bigr](\Gamma) \, \re^{N^2 F_0 +F_1} \Theta_{ \mu, \nu} (u,\tau)
\eeq
under a general ${\rm Sp}(2\bar g, \IZ)$ transformation. This proves (\ref{tlaw}) to leading order in $N$.

 \begin{figure}[!ht]
\leavevmode
\begin{center}
\epsfxsize=12cm
\epsfysize=4.5cm
\epsfbox{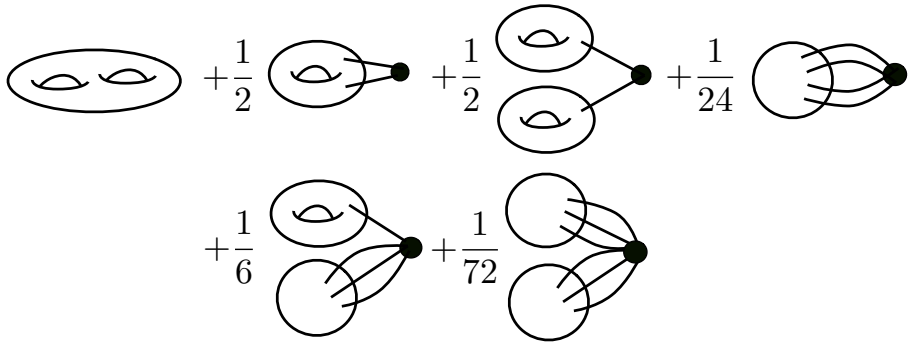}
\end{center}
\caption{Graphic representation of $Z_2$, the term in $1/N^2$ in the expansion of $Z$.}
\label{z2}
\end{figure}

\medskip

Our goal is now to prove that each $Z_j$ is itself modular invariant, up to the change of characteristics (\ref{charchange}).

$Z_j$ is given by a sum,  and each term in this sum is the product of a derivative of $\Theta$, with a product of a certain number of derivatives of $F_h$'s corresponding to the stable degeneracy of (possibly non-connected) surfaces with Euler characteristics $\chi=-j$. Notice that, in each term, the total number of derivatives acting on the $F_h$ equals the number of derivatives acting on $\Theta$. As shown in \figref{fig:modulardic}, 
$\Theta^{(n)}/\Theta$, the $n^{\rm th}$ derivative of $\Theta$, is depicted as a black dot with $n$ legs, while 
$F_h^{(n)}$ is depicted as a Riemann surface of genus $h$ with $n$ marked points. Each term in $Z_j$ can thus be depicted as a set of stable Riemann surfaces with marked points linked to the legs of the $\Theta^{(n)}/\Theta$ term. Notice that there can be disconnected terms corresponding to $F_h^{(n)}$ with $n=0$.
Also, each term has a symmetry factor, which is, as usual in Feynmann graphs, the number of possible pairings of indices, divided by the group order $k!\, l_1!\dots l_k!$.
As an example, we depict $Z_2$ in \figref{z2}.



 \begin{figure}[!ht]
\leavevmode
\begin{center}
\epsfxsize=10cm
\epsfysize=3cm
\epsfbox{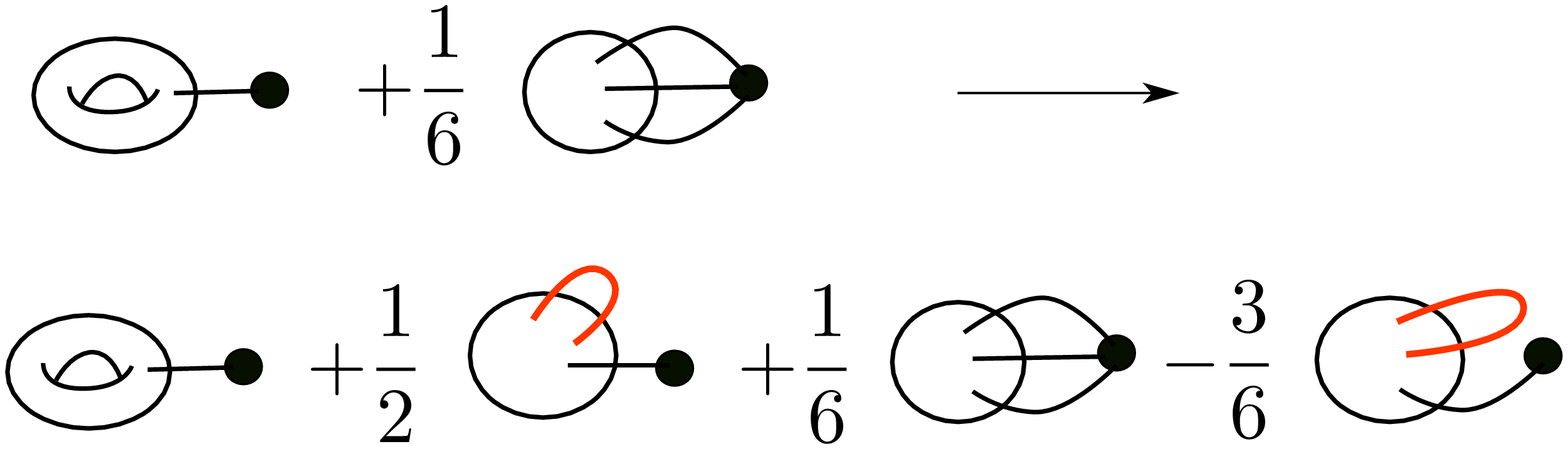}
\end{center}
\caption{Graphic proof of the modular invariance of the term $Z_1$ of order $1/N$ in the expansion of $Z$.}
\label{fig:modular3}
\end{figure}

Under a modular transformation, each $F_h^{(n)}$ transforms into a sum of stable degenerated surfaces, with a factor $\kappa/(2\pi \ri)$ for each degenerated cycle. As shown in (\ref{thetadergen}) (and as illustrated in \figref{fig:thetader}) the $\Theta^{(n)}/\Theta$ term transforms into a sum of $\Theta^{(k)}/\Theta$ where $(n-k)/2$ legs are replaced by a $-\kappa/(2\pi\ri )$ in all possible ways (with a change of characteristics as (\ref{charchange}). The factors of $C\tau + D$ appearing in (\ref{thetadergen}) cancel against the factors of $(C\tau +D)^{-1}$ coming from the derivatives of the $F_h$'s.

 \begin{figure}[!ht]
\leavevmode
\begin{center}
\epsfxsize=12cm
\epsfysize=4.5cm
\epsfbox{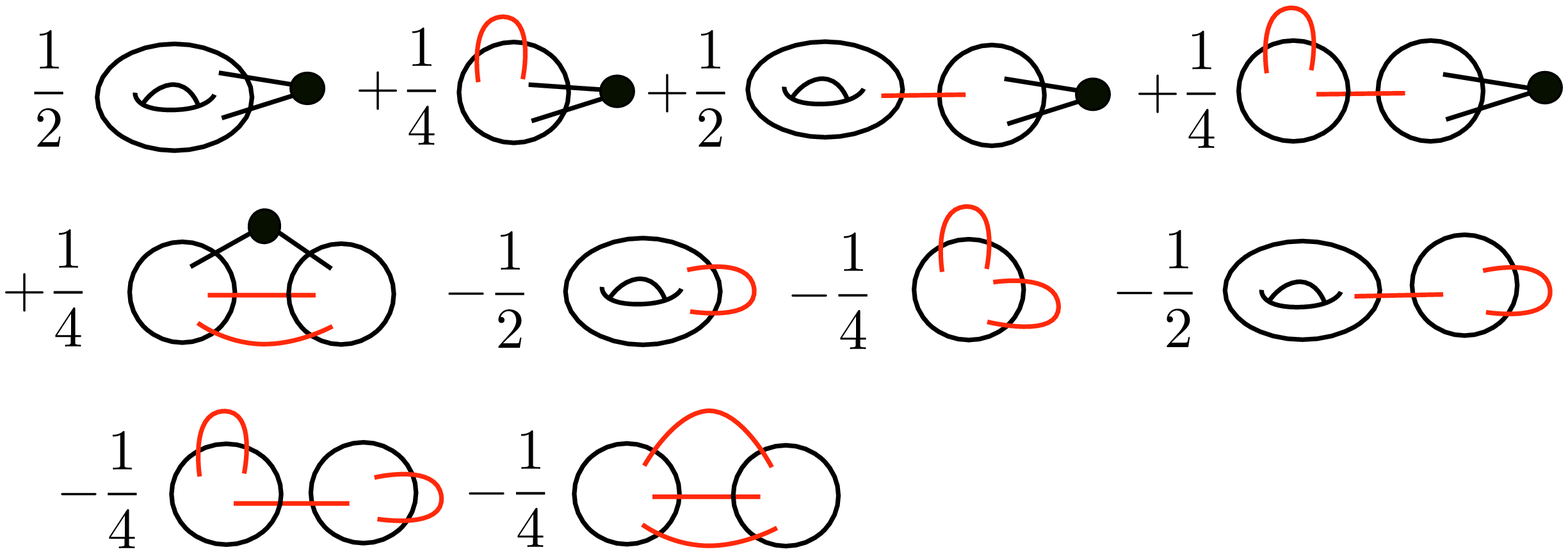}
\vskip.8cm
\epsfxsize=11cm
\epsfysize=4.5cm
\epsfbox{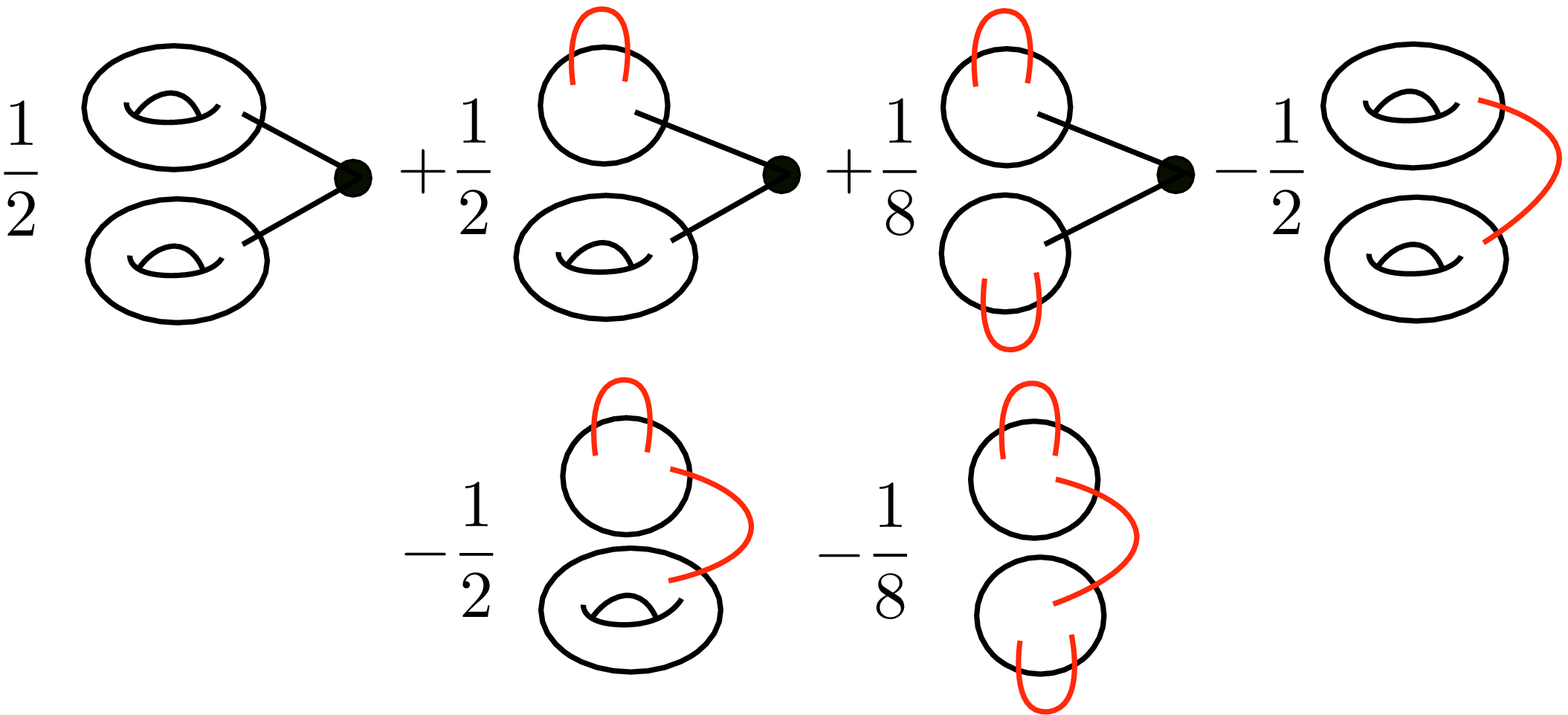}
\vskip.5cm
\epsfxsize=11cm
\epsfysize=4.5cm
\epsfbox{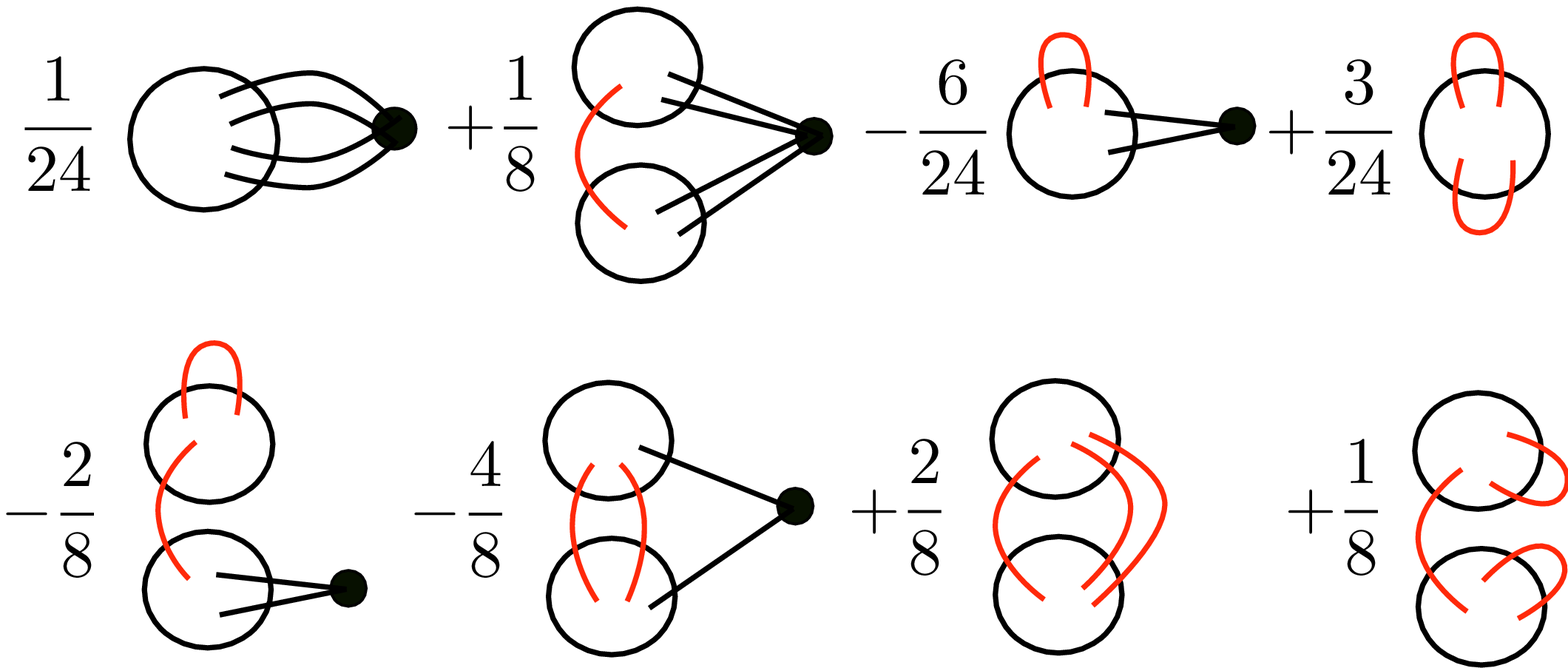}
\end{center}
\caption{Modular transformation of the second, third and fourth graphs in \figref{z2} .}
\label{n2one}
\end{figure}

\bigskip
In the end, $\td{Z}_j$ is a sum of terms, each of them is the product of  a $\Theta^{(k)}/\Theta$, with a certain number of $F_{h_i}^{(n_i)}$'s, and a certain number of $\kappa$ propagators. Each propagator can be obtained in two ways, either from a degeneracy of a $F_h^{(n)}$, with a factor $+(2 \ri\pi)^{-1}$, or from a term $\Theta^{(k)}/\Theta$, with a factor $-(2 \ri\pi)^{-1}$.
Also, the symmetry factors are counting the number of possible pairings corresponding to a given diagramm, and are the same whether obtained from the modular transformation of $\Theta^{(k)}/\Theta$, or from the modular transformation of $F_{h_i}^{(n_i)}$'s.
Therefore, the total contribution of each $\kappa$ propagator is $0$. This shows that
\beq
\td{Z}_j (\tilde \mu, \tilde \nu) =Z_j (\mu, \nu)
\eeq
therefore we have proved that 
\be
\tilde Z_{\Sigma}(\tilde \mu,\tilde \nu;\tilde \epsilon)=\zeta\bigl[^\mu_\nu\bigr](\Gamma)\,\, Z_{\Sigma}(\mu, \nu;\epsilon)
\eeq

We illustrate the proof by the examples of $Z_{1}$ and $Z_2$. In the case of $Z_1$, we know from \figref{genusoneder} the transformation 
rule of $F_1'$, and from \figref{fig:thetader} we know the transformation properties of $\Theta'/\Theta$ and $\Theta'''/\Theta$. Since 
$F_0'''$ transforms as a modular form of weight $-3$ (i.e., there are no shifts involved) and it is completely symmetric in its indices, 
we immediately obtain the graphic proof of modular invariance of $Z_1$ depicted in \figref{fig:modular3}.

 \begin{figure}[!ht]
\leavevmode
\begin{center}
\epsfxsize=11cm
\epsfysize=4.5cm
\epsfbox{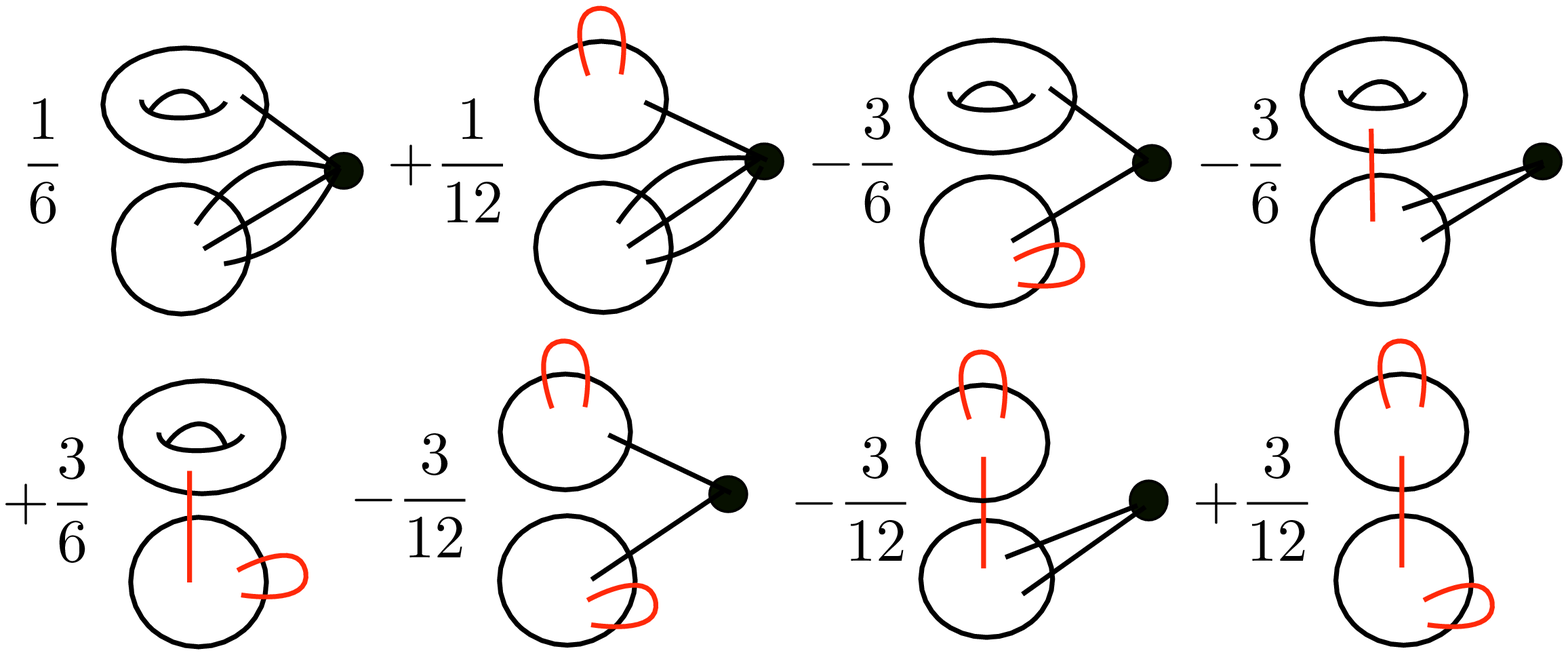}
\vskip .8cm
\epsfxsize=11cm
\epsfysize=4.5cm
\epsfbox{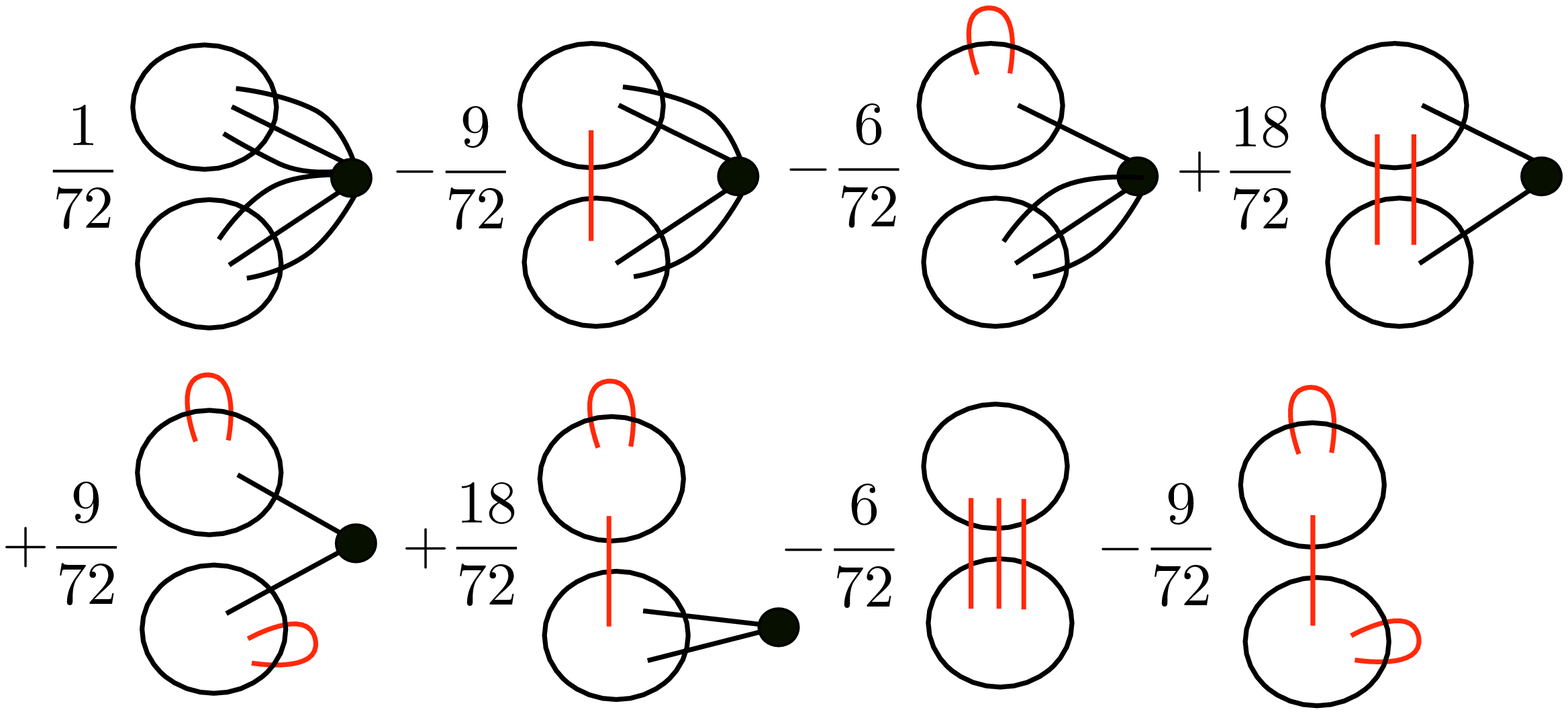}
\end{center}
\caption{Modular transformation of the fifth and sixth graphs in \figref{z2} .}
\label{n2two}
\end{figure}

The terms appearing in $Z_2$ were depicted graphically in \figref{z2}. The modular transformation of $F_2$, which is the first term, was shown in \figref{f2modular}. The modular transformation of the rest of the graphs in \figref{z2} is shown in \figref{n2one} and \figref{n2two}. One can check that 
all the graphs generated on top of the original ones (i.e., all the shifts generated by the modular transformation, i.e. all graphs containing propagators $\kappa$) eventually cancel, and $Z_2$ remains invariant up to the change of characteristics. For example, the second graph in the first line of \figref{n2one} cancels against the third graph in the 
sixth line of the same figure, while the third graph in the first line of \figref{n2one} cancels against the fourth graph in the first line of \figref{n2two}. In a similar manner, all graphs in \figref{f2modular}, 
\figref{n2one} and \figref{n2two} containing propagators eventually cancel.

\subsection{Background independence}

We now show that the non-perturbative partition function is background independent, i.e. it does not depend on the background filling fraction $\epsilon$. This is not surprising, because this is how it was first constructed in \cite{eone}.

\smallskip
Let
\beq
\hat{Z}(\eta) = \ee{\sum_{g\geq 0} N^{2-2g} F_g(\eta)} 
\eeq
be the perturbative partition function. We may Taylor expand it near an arbitrary background filling fraction $\epsilon$:
\beq
\ba \hat{Z}(\eta) 
&=  \sum_j {1\over j!}(\eta-\epsilon)^j \,\, \hat{Z}^{(j)}(\epsilon) \\
&= \ee{N^2(F_0 + (\eta-\epsilon)F'_0 + {1\over 2} (\eta-\epsilon)^2 F''_0)}\, \ee{F_1}\, \, \ee{\sum_{g\geq 2} N^{2-2g} F_g}\\
& \cdot \sum_{k} \sum_{l_i>0}\sum_{h_i>1-{l_i\over 2}} {N^{\sum_i (2-2h_i)}\over k! l_1!\,\dots\, l_k!} {F_{h_1}^{(l_1)}}\dots {F_{h_k}^{(l_k)}}  \,\, (\eta-\epsilon)^{\sum_i l_i}.\\
\ea
\eeq
Therefore we recognize that the nonperturbative partition function (\ref{exphatZTheta}) is the Taylor expansion of
\be
Z_{\Sigma}(\mu,\nu;\epsilon)
= \sum_{{\bf n}\in {\mathbb Z}^{\bar g}} 
\hat{Z}\Bigl({n+\mu\over N}\Bigr)\,\, \ee{2 \ri\pi n \nu}.
\eeq
The right hand side of this expression is clearly independent of $\epsilon$:
\beq
Z_{\Sigma}(\mu,\nu;\epsilon) = Z_{\Sigma}(\mu,\nu;\epsilon')
\eeq
and thus $Z_{\Sigma}(\mu,\nu;\epsilon)=Z_{\Sigma}(\mu,\nu)$ is background independent.

\medskip

We can see more explicitly that $Z_{\Sigma}(\mu,\nu;\epsilon)$ is locally constant, as a function of $\epsilon$, by showing that $\rd Z/\rd \epsilon=0$. 
Let us first define
\beq
\label{thetak}
\Theta_k := \ee{N^2 F_0+F_1}\,\, \left.{\rd ^k\over \rd u^k} \Theta\right|_{u=NF'_0}.
\eeq
In this equation and the following ones we will not indicate the characteristics of the theta function in order to simplify the notation. 
Using (\ref{thetak}) we can write
\be
\ba
Z_{\Sigma}(\mu,\nu;\epsilon) 
&= \Theta_0 + {1\over N}\Bigl(F'_1 \Theta_1+{F_0'''\over 6}\Theta_3\Bigr) \\
& + {1\over N^2}\Bigl(F_2 \Theta_0+{F_1''\over 2}\Theta_2+{F_1'^2\over 2}\Theta_2+{F_0''''\over 24}\Theta_4+{F_0'''F_1'\over 6}\Theta_4+{F_0'''^2\over 72}\Theta_6\Bigr)+\dots 
\ea
\eeq
It is easy to compute that (where $'=\rd/\rd\epsilon$)
\beq
\Theta_k' = F'_1 \Theta_k - kN \Theta_{k-1} + {F_0'''\over 2}\Theta_{k+2}
\eeq
and, as a formal power series, we have that
\be
\ba
{\rd Z_{\Sigma} \over \rd \epsilon}
&= \Theta_0' + {1\over N}\Bigl(F''_1 \Theta_1 +F'_1 \Theta'_1 + {F_0''''\over 6}\Theta_3+{F_0'''\over 6}\Theta_3'\Bigr) \\
&  + {1\over N^2}\Bigl(F_2' \Theta_0+F_2 \Theta_0'+{F_1'''\over 2}\Theta_2+{F_1''\over 2}\Theta_2'+{F_1'^2\over 2}\Theta_2'+F_1'F_1''\Theta_2+{F_0^{(5)}\over 24}\Theta_4 \\
& +{F_0''''\over 24}\Theta_4'+{F_0'''F_1''\over 6}\Theta_4+{F_0''''F_1'\over 6}\Theta_4+{F_0'''F_1'\over 6}\Theta_4'+{F_0''' F_0''''\over 36}\Theta_6+{F_0'''^2\over 72}\Theta_6'\Bigr)+\cdots \\
&= F'_1 \Theta_0 + {F_0'''\over 2}\Theta_2 + {1\over N}\Bigl(F''_1 \Theta_1 +F'_1(F'_1\Theta_1-N\Theta_0+{F_0'''\over 2}\Theta_3) + {F_0''''\over 6}\Theta_3 \\
& +{F_0'''\over 6}(F'_1\Theta_3-3N\Theta_2+{F_0'''\over 2}\Theta_5)\Bigr) \\
&  + {1\over N^2}\Bigl(F_2' \Theta_0+F_2 (F'_1\Theta_0+{F_0'''\over 2}\Theta_2)+{F_1'''\over 2}\Theta_2+{F_1''\over 2}(F'_1\Theta_2-2N\Theta_1+{F_0'''\over 2}\Theta_4) \\
& +{F_1'^2\over 2}(F'_1\Theta_2-2N\Theta_1+{F_0'''\over 2}\Theta_4)+F_1'F_1''\Theta_2+{F_0^{(5)}\over 24}\Theta_4 \\
& +{F_0''''\over 24}(F'_1\Theta_4-4N\Theta_3+{F_0'''\over 2}\Theta_6) \\
& +{F_0'''F_1''\over 6}\Theta_4+{F_0''''F_1'\over 6}\Theta_4+{F_0'''F_1'\over 6}(F'_1\Theta_4-4N\Theta_3+{F_0'''\over 2}\Theta_6) \\
& +{F_0''' F_0''''\over 36}\Theta_6+{F_0'''^2\over 72}(F'_1\Theta_6-6N\Theta_5+{F_0'''\over 2}\Theta_8)\Bigr)+\cdots \\
\ea
\eeq
One easily checks that the term of order $N^0$ in this series is
\beq
F'_1 \Theta_0 + {F_0'''\over 2}\Theta_2 + {1\over N}(F'_1 (-N \Theta_0) + {F_0'''\over 6} (-3N\Theta_2) ) =0.
\eeq
The term of order $1/N$ is
\be
\ba
&F''_1 \Theta_1 +F'_1(F'_1\Theta_1+{F_0'''\over 2}\Theta_3) + {F_0''''\over 6}\Theta_3+{F_0'''\over 6}(F'_1\Theta_3+{F_0'''\over 2}\Theta_5) \\
&  + {1\over N}\Bigl({F_1''\over 2}(-2N\Theta_1)  +{F_1'^2\over 2}(-2N\Theta_1)+{F_0''''\over 24}(-4N\Theta_3) +{F_0'''F_1'\over 6}(-4N\Theta_3) +{F_0'''^2\over 72}(-6N\Theta_5)\Bigr) \\
&  =0.
\ea
\eeq
This shows that, at this order in the $1/N$ expansion, $Z_{\Sigma}(\mu,\nu;\epsilon)$ is locally constant as a function of the filling fractions:
\beq
{\rd Z_{\Sigma} \over \rd \epsilon}=0.
\eeq
The generalization at all orders can be done by using the graphical techniques developed above. 

\section{Matrix models and sums over filling fractions}
\label{sectionMM}

Let us briefly summarize \cite{eone}, which is just an improved version of \cite{bde}, to explain the origin of the definition (\ref{exphatZTheta}) for the non-perturbative partition function.

\subsection{Matrix integrals and domains of integration}

A matrix integral is typically an integral of the form:
\beq\label{typZRMT}
Z = {1\over N!}\,\int \rd x_1\dots \rd x_N\,\, \prod_{i>j} (x_i-x_j)^2\,\, \prod_i \ee{-NV(x_i)}
\eeq
where we have not written the integration domain.
For any choice of integration domain, $Z$ satisfies the same loop equations (i.e. Schwinger-Dyson equations), as long as there is no boundary term when one integrates by parts.

For example, one may consider the matrix integral over $H_N(\gamma)$ (the set of normal matrices with eigenvalues constrained on a path $\gamma$):
\beq
Z(\gamma) = \int_{H_N(\gamma)} \rd M\,\ee{-N\Tr V(M)}= {1\over N!}\,\int_{\gamma^N} \rd x_1\dots \rd x_N\,\, \prod_{i>j} (x_i-x_j)^2\,\, \prod_i \ee{-NV(x_i)}.
\eeq
This integral depends only on a choice of path $\gamma$, and it has a $U(N)$ invariance.

\smallskip
One may also consider the following integral where $N=\sum_i n_i$:
\beq
\hat{Z}_{n_1,\dots,n_k} = {1\over \prod_i n_i!}\,\int_{\gamma_1^{n_1}\times\dots\times \gamma_k^{n_k}} \rd x_1\dots \rd x_N\,\, \prod_{i>j} (x_i-x_j)^2\,\, \prod_i \ee{-NV(x_i)}.
\eeq
The parameters $\epsilon_i=n_i/N$ are called the filling fractions, because they represent the fraction of the eigenvalues on each path $\gamma_i$.
This integral depends on a choice of filling fractions, called ``background filling fraction", and it only has a $U(n_1)\times U(n_2)\times \dots \times U(n_k)$ invariance.

\medskip

One may also write a path in different ways, and for instance if $\gamma_1,\dots,\gamma_k$ form a homological basis of paths on which integrals of type (\ref{typZRMT}) can be performed \cite{Marcopaths}, then any path $\gamma$ can be decomposed on such a basis:
\beq\label{gammacigammai}
\gamma=\sum_i c_i \gamma_i
\eeq
and one may relate $Z(\gamma)$ and $\hat{Z}_{n_1,\dots,n_k}$:
\beq\label{ZgammaZhat}
Z(\gamma)  = \sum_{\sum_i n_i=N}\, \prod_i c_i^{n_i}\,\, \hat{Z}_{n_1,\dots,n_k}.
\eeq
Since not all $n_i$ are independent, we can fix the overall normalization of $Z(\gamma)$ by setting $c_1=1$. One may also study changes of paths in $\hat{Z}_{n_1,\dots,n_k}$.

\subsection{Topological perturbative expansion}

For some special choices of $\gamma$, or alternatively for some special choices of the basis $\gamma_1,\dots,\gamma_k$ and some special choices of filling fractions $n_1,\dots,n_k$, it may happen that $\ln{Z}$, or $\ln{\hat{Z}}$ have a perturbative large $N$ expansion:
\beq\label{hatZnipertexp}
\ln{\hat{Z}_{n_1,\dots,n_k}} = \sum_{g=0}^\infty N^{2-2g} \, F_g.
\eeq
If such $F_g$'s exist, then Schwinger-Dyson equations imply \cite{eynloopeq, ce} that they coincide with the symplectic invariants defined in \cite{eo}, for some spectral curve. This matrix model spectral curve is the so called ``equilibrium density" of eigenvalues.

\medskip
However,  for most paths $\gamma$, or for arbitrary choices of the basis $\gamma_1,\dots,\gamma_k$ and arbitrary choices of filling fractions $n_1,\dots,n_k$, matrix integrals have no perturbative large $N$ expansion.

\medskip

We conjecture (which is proved for special cases, for instance for the one-matrix model \cite{bertola}), that given a potential $V$, and given $n_1,\dots,n_k$, there always exists a ``good" basis of paths $\gamma_1,\dots,\gamma_k$, such that $\hat{Z}_{n_1,\dots,n_k}$ has a topological expansion of the form (\ref{hatZnipertexp}).
In some sense, this ``good" basis of paths is made of the steepest descent contours for the integral (\ref{typZRMT}).

\subsection{Summation over filling fractions}

Using (\ref{ZgammaZhat})
\beq\label{Zgammasumff}
Z(\gamma)  = \sum_{\sum_i n_i=N}\, \prod_i c_i^{n_i}\,\, \hat{Z}_{n_1,\dots,n_k}
\eeq
and the fact that each $\hat{Z}_{n_1,\dots,n_k}$ has a perturbative expansion of type (\ref{hatZnipertexp}), we can find the asymptotic expansion of $Z(\gamma)$ as a combination of $\hat{Z}_{n_1,\dots,n_k}$ after summation over the filling fractions $n_i$. This goes as follows. 

\begin{figure}[!ht]
\leavevmode
\begin{center}
\epsfxsize=12cm
\epsfysize=5cm
\epsfbox{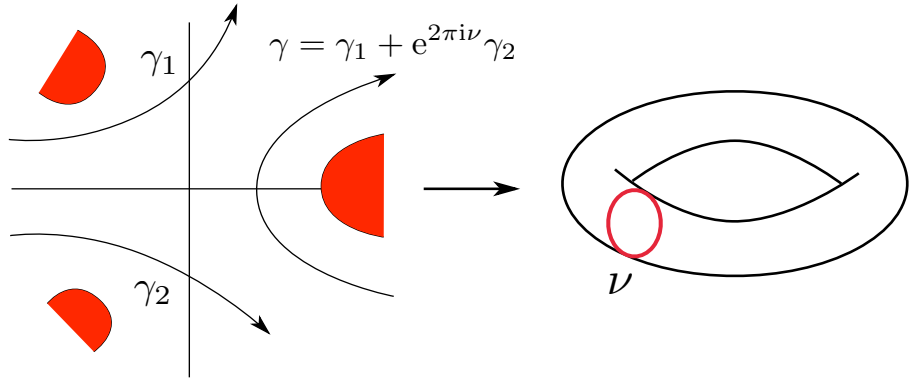}
\end{center}
\caption{A choice of integration path for the matrix integral determines a choice 
of characteristic in the spectral curve, corresponding to A-cycles. We illustrate this correspondence in this figure for the case of the cubic matrix model, with a spectral curve of genus one. A typical path of integration for the eigenvalues is $\gamma=\gamma_1+\ee{2\ri\pi\nu}\gamma_2$ and corresponds to a characteristic $(0,\nu)$.}
\label{fig:nppf}
\end{figure}

For various matrix models, including the 1-matrix model, the 2-matrix model, the chain of matrices, and matrix models with external fields, the coefficients $F_g$ in (\ref{hatZnipertexp}) have been computed in \cite{ce, ceo, ep, eo}, and are the symplectic invariants of \cite{eo}. They happen to be analytic in the filling fraction variables $n_i/N$, and thus they can be Taylor expanded near an arbitrary background value $\epsilon_i$:
\beq
N^{2-2g}\,F_g(n_i/N) = \sum_k {N^{2-2g-k}\,(n_i-N\epsilon_i)^k\over k!}\, F_g^{(k)}(\epsilon_i).
\eeq
The only terms with non-negative powers of $N$ correspond to $N^2 F_0 + F_1 + (n-N\epsilon) NF'_0 + {1\over 2}(n-N\epsilon)^2F_0''$.
All the other terms have negative power of $N$, and thus can be expanded at large $N$, so that:
\beq
\ba
\hat{Z}_{n_1,\dots,n_k}
& \sim \ee{N^2 F_0 + F_1+\sum_{g\geq 2} N^{2-2g} F_g}\,\, \\
& \sum_k 
\sum_{l_i>0}\sum_{h_i>1-{l_i\over 2}} {N^{\sum_i (2-2h_i-l_i)}\over k! l_1!\,\dots\, l_k!}\,\,\, F_{h_1}^{(l_1)}\dots F_{h_k}^{(l_k)}  \,\, (n-N\epsilon)^{\sum_i l_i}\,\, \ee{(n-N\epsilon) NF'_0 + {1\over 2}(n-N\epsilon)^2F_0''}. \\
\ea
\eeq
The sum over filling fractions (\ref{Zgammasumff}) 
generates the non-perturbative partition function of \cite{eone}, i.e. (\ref{exphatZTheta}), under the identification:
\beq
\mu_i = 0, \qquad c_i = \ee{2\ri\pi \nu_i}.
\eeq
This shows that the choice of a characteristic $(\mu,\nu)$ is related to a choice of integration contour $\gamma$ for the matrix integral (see \figref{fig:nppf}).

\smallskip

This simplified description of the origin of the non-perturbative partition function for matrix models seems to involve only characteristics of type $(0,\nu)$, and not all characteristics $(\mu,\nu)$.

\begin{figure}[!ht]
\leavevmode
\begin{center}
   \epsfxsize=0.5\textwidth
    \leavevmode
    \mbox{\hspace{-.5cm}\epsfxsize=5cm
\epsfysize=5.5cm\epsfbox{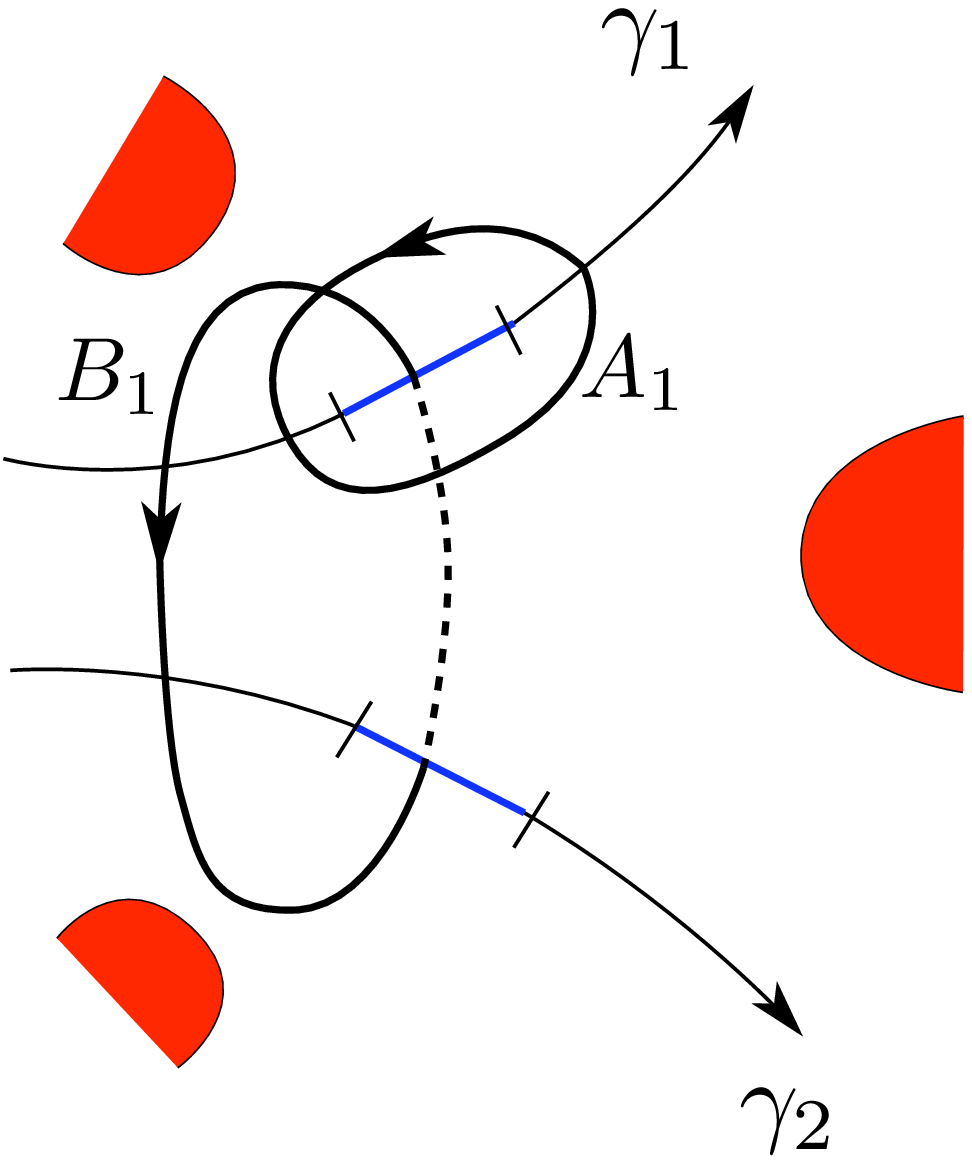}\qquad\quad\epsfxsize=7cm
\epsfysize=5cm\epsfbox{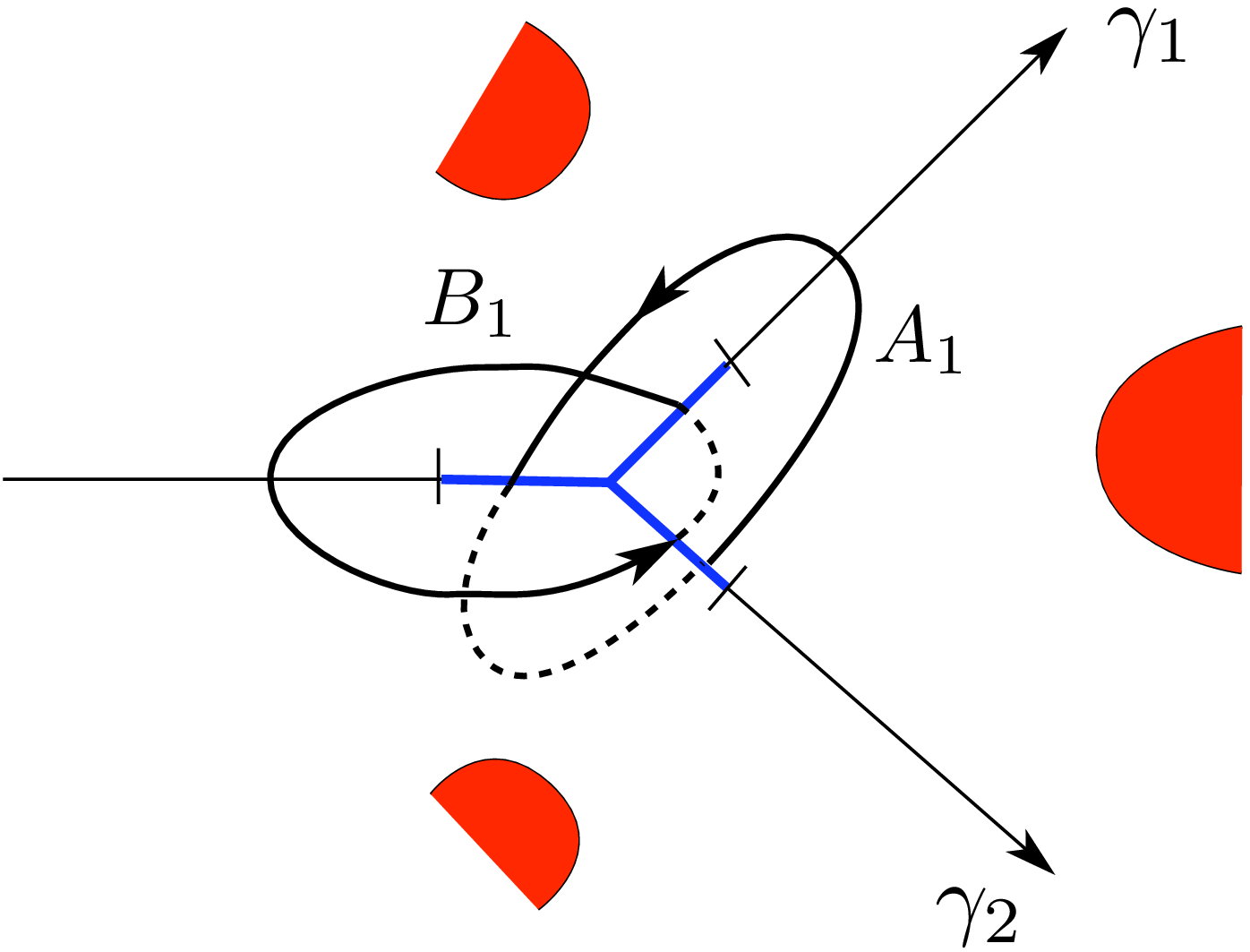}\hspace{-.5cm}}
    \end{center}
    \caption{On the left we show a situation in which there is a natural choice of $A$-cycle surrounding the cut along the path $\gamma_1$. However, in some situations the eigenvalues might localize on tree structures, as shown in the figure on the right.}
\label{contours}
\end{figure}

However, one should keep in mind that a ``good basis of paths" $\gamma_1,\dots,\gamma_k$ is not unique. 
Changes of basis correspond to modular transformations, and basis can be more complicated than what naive intuition seems to show.
In the ``naive" picture, it is often thought that, in the large $N$ limit of mutlticut matrix integrals, eigenvalues tend to localize along disconnected segments called ``cuts".
In this case, there is a natural choice of $A$-cycles, as contours surrounding the cuts (see \figref{contours}, left). 

But in general, it is well known \cite{david, bertola} that, in the large $N$ limit of multicut matrix integrals, eigenvalues tend to localize along forests of 3-valent trees rather than union of segments (see \figref{contours}, right).
In that case, there is no natural distinction between $A$-cycles and $B$-cycles, and maybe in such tree structures one could probably interpret the general characteristics $(\mu,\nu)$ with $\mu\neq 0$.

\subsection{Remarks on background independence}

First, notice that we have chosen to perform a Taylor expansion near an arbitrary background filling fraction $\epsilon$, although the partition function $Z(\gamma)$ is independent of $\epsilon$. We clearly have background independence.


\medskip

Also, we have used the existence of a good basis $\gamma_1,\dots,\gamma_k$. However, such a good basis is not unique, and one may perform changes of basis.
Such changes of basis are equivalent to symplectic changes of cycles on the spectral curve.
The fact that $Z(\gamma)$ depends intrinsically on $\gamma$, and not on a choice of basis on which $\gamma$ is decomposed as in (\ref{gammacigammai}), is a hint towards modular invariance.

Finally, we notice that a choice of background leads to a breaking of the original, unitary $U(N)$ symmetry of 
the matrix integral down to a subgroup $U(n_1) \times \cdots \times U(n_k)$. 
Therefore, the restoration of modular invariance and background 
independence by non-perturbative corrections seems to be deeply related to the restoration of the $U(N)$ symmetry.

\sectiono{The nonperturbative partition function for topological strings}

\subsection{Topological strings, matrix models and spectral curves}

 In \cite{dv}, Dijkgraaf and Vafa showed that the type B topological string, on certain Calabi-Yau  (CY) geometries, is 
 equivalent to a Hermitian 
 matrix model with polynomial potential $V(x)$ (see \cite{mmhouches} for a review). These target geometries are deformed singularities of the form
\be\label{scts}
uv=H(x,y)
\eeq
where $x,y,u,v \in \IC$ and 
\be
H(x,y)=y^2-V'^2(x)- f(x). 
\eeq
The nontrivial information about this geometry turns out to be encoded in the Riemann surface $\Sigma$ described by 
$H(x,y)=0$. An important insight of the analysis of Dijkgraaf and Vafa is that $\Sigma$ is precisely the spectral curve of the corresponding Hermitian matrix model, providing in this way a beautiful example in which the master field of the $1/N$ expansion generates the target geometry of a string theory.

More recently, the correspondence between matrix models/spectral curves 
and topological strings was extended to toric CY threefolds \cite{mm,bkmp}. 
This class of examples is very interesting since (in contrast to the geometries considered in \cite{dv}) they have mirror geometries. 
The mirrors are described by an equation of the form (\ref{scts}), but where the variables $x,y$ now belong to $\IC^*$, and again the nontrivial part of the geometry reduces to a Riemann surface $\Sigma$ described by $H(x,y)=0$. In \cite{eo} it was shown that, given any algebraic curve, 
not necessarily coming from a matrix model, one can compute a set of ``open" and ``closed" amplitudes. 
When the curve is the spectral curve of a matrix model, these amplitudes reproduce the $1/N$ expansion of 
correlators and free energies, respectively. However, the construction is valid for any spectral curve. The proposal of \cite{mm,bkmp} is to regard the 
Riemann surface $H(x,y)=0$ appearing in (\ref{scts}) as a spectral curve in the general sense of \cite{eo}, 
and to identify the open and closed string amplitudes of the B--model with the open and closed amplitudes constructed in 
\cite{eo} purely in terms of data of the spectral curves. Therefore, for this more general class of examples, 
we rather have a correspondence between the type B topological string amplitudes on the mirror of a toric manifold, 
and the amplitudes associated to a spectral curve defined in \cite{eo}. 

For certain backgrounds it is however still possible to provide a matrix integral representation. In the case of local curves, i.e. manifolds 
of the form 
\be
X_p =\CO(p) \oplus \CO(-p-2) \rightarrow \IP^1
\eeq
this representation can be obtained directly from the A-model answer for the amplitudes, written as a sum over partitions \cite{epartitions}. Another family of Calabi--Yau's where a matrix integral representation is available is 
given by
\be
\label{localXp}
Y_p=S_{A_{p-1}} \rightarrow \IP^1,
\eeq
where $S_{A_{p-1}}$ is a four-dimensional $A_{p-1}$ resolved singularity. For example, for $p=2$, $S_{A_1}$ is a resolved $\IC^2/\IZ_2$ singularity and 
we can represent it also as 
\be
\label{localX2}
Y_2= K^{-1}  \rightarrow \IP^1 \times \IP^1, 
\eeq
i.e. as the total space of the anticanonical bundle of $\IP^1\times \IP^1$. These backgrounds have a large $N$ dual \cite{akmv} given by 
Chern--Simons theory with gauge group $U(N)$ on the lens space
\be
L(p,1)=\IS^3/\IZ_p.
\eeq
 For $p=1$ this is just the Gopakumar--Vafa duality of \cite{gv} (extensions to $L(p,q)$ have been 
recently discussed in \cite{andreaetal}). 
It turns out that Chern--Simons theory on $L(p,1)$ has a matrix integral realization \cite{mmcsp}, and it can be shown that the spectral curve of this matrix integral reproduces the mirror geometry \cite{hy}. Therefore, for those backgrounds, one can indeed give a nonperturbative definition in terms of a matrix model. 

\subsection{Topological strings and the nonperturbative partition function}

It has been argued in \cite{msw,mmnp} that the above correspondence between matrix models/spectral curves and topological strings could also be used to obtain nonperturbative information on the topological string side. In particular, instanton effects on the matrix model side should lead to spacetime instantons 
in topological string theory, generalizing in this way the connection between eigenvalue tunneling and ZZ branes in noncritical string theory \cite{martinec,akk}. This has been verified in some simple models in \cite{msw,mmnp} by checking that these   
instanton effects control the large order behavior of string perturbation theory. 

Since the nonperturbative partition function of \cite{eone} studied in this paper incorporates the full multi-instanton series of the matrix model and has good modular properties, it is natural to generalize \cite{msw,mmnp} and use $Z_{\Sigma}(\mu,\nu)$ to define topological string theory nonperturbatively. Notice that the ingredients to construct $Z_{\Sigma}(\mu, \nu)$ are simply the spectral curve and a choice of characteristics. All the nonperturbative information is encoded in the latter, since they correspond to (and generalize) the choice of contour in the matrix integral. 

We then make the following proposal. Let $X_{\Sigma}$ be a local Calabi--Yau manifold whose geometry is encoded in a spectral curve 
$\Sigma$. For each choice of characteristics $\mu, \nu$ on $\Sigma$, we define the nonperturbative topological string partition function $Z_{X_{\Sigma}} (\mu, \nu)$ associated to the characteristics $\mu,\nu$ as
\be
\label{proposal}
Z_{X_{\Sigma}} (\mu, \nu)=Z_{\Sigma}(\mu,\nu).
\eeq
In this equality, the filling fractions are identified with the moduli of the Calabi--Yau as in \cite{dv,akmv}, and the string coupling constant is identified with $1/N$: 
\be
 t^i=\epsilon^i, \qquad g_s={1\over N}.  
 \eeq
In principle, each choice of characteristic provides a nonperturbative completion of 
topological string theory, as it was argued in \cite{mmnp} in some simple examples. This situation is of course 
reminiscent of the nonperturbative ambiguity in 2d gravity, where different choices of integration contour provide different 
nonperturbative completions. 

Our proposal (\ref{proposal}) incorporates previous suggestions concerning the 
nonperturbative structure of topological string theory. In \cite{mmss} 
it was pointed out, based on results for noncritical strings, that the exact topological string partition function should be a sum over all vacua, i.e. over all instantons obtained by filling all critical points of the potential. This is precisely what was done in \cite{eone} to obtain $Z_{\Sigma}(\mu,\nu)$, and in fact it is at the origin of its background independence. $Z_{X_\Sigma}(\nu, \mu)$ is also closely related to the generalized partition functions introduced in section 4.7 of \cite{adkmv} and in section 3.4 of \cite{dhsv}, in the context of a fermion formulation of topological string theory on the spectral curve $\Sigma$. In both cases, one considers a partition function where one sums over all possible fermion numbers with fixed twists around the A cycles, i.e. one goes to a grand canonical ensemble for the fermions where the twists correspond to chemical potentials. The relation to a fermion system is now further clarified, since what we have seen in this paper is that $Z_{\Sigma}(\mu,\nu)$ has the same modular properties as the partition function of a twisted fermion on $\Sigma$, with twist $\mu,\nu$. We will extract more consequences of this below.

Although our proposal does not 
completely solve the problem of providing a nonperturbative definition of topological string theory, even in the 
local case, it cuts down the freedom in making such a definition to a well--defined choice: a characteristic on the Riemann surface. Moreover, 
there are some nonperturbative effects that do depend on the multi-instanton amplitudes encoded in 
(\ref{proposal}) but not on the choice of characteristic. One example is the large order 
behaviour of string perturbation theory, as discussed in \cite{msw,mmnp}.  

How reasonable is our proposal (\ref{proposal})? As it was pointed out in \cite{mmnp}, for the Calabi--Yau's of the form (\ref{localXp}), 
the topological string theory has a 
large $N$ Chern--Simons dual, and the total partition function is indeed of the form $Z_{\Sigma}(\mu,\nu)$. Moreover the characteristics, i.e. the nonperturbative data, are fixed. To see this, one notices that at finite $N$ one has \cite{mmcsp,akmv,mmnp}
\be
\label{csmatrix}
Z_{\rm CS}(N, g_s) =\sum_{N_1+\cdots+N_p=N} c_1^{N_1} \cdots c_p^{N_p} Z(N_1, \cdots, N_p), 
\eeq
where 
\be\label{charcs}
c_j=\exp\Bigl( {\pi \ri \hat k \over p} (j-1)^2 \Bigr), \quad j=1, \cdots, p
\eeq
and $\hat k=k+N$ is the shifted coupling constant of Chern--Simons theory, which is related to the string coupling constant by 
\be
g_s= {2\pi \ri \over p \hat k}. 
\eeq
The r.h.s of (\ref{csmatrix}) is of the form (\ref{Zgammasumff}), therefore at large $N$ is nothing but $Z_{\Sigma_p} (0,\nu)$, where the spectral curve $\Sigma_p$ is the Riemann surface appearing 
in the mirror of $Y_p$ and it has genus $\bar g=p-1$. Since $c_1$ is already normalized to one, we read that the characteristic $\nu$ is given by 
\be
\nu_j={\hat k \over 2 p} j^2, \quad j=1, \cdots, \bar g
\eeq
while $\mu=0$. Notice that, in Chern--Simons theory, $\hat k$ is an integer, and the $c_j$ are indeed phases which depend 
on truly nonperturbative data, i.e. the value of $k+N$ as a quantized integer (the quantization of $k$ in Chern--Simons theory 
is not visible in perturbation theory, nor in the large $N$ expansion). Therefore, 
for these backgrounds, the partition function of the holographic Chern--Simons dual is precisely the nonperturbative partition function $Z_{\Sigma_p}(\mu,\nu)$ with a definite choice of characteristics. The proposal (\ref{proposal}) is then completely natural in this case.   

If we now consider a general local Calabi--Yau background, is there any way of 
fixing the characteristics, i.e. the nonperturbative content of the theory? A possible guiding principle is the following. Note 
that, if $\mu_i, \nu_i \in \IR$, the partition function $Z_{\Sigma} (\mu,\nu)$ 
transforms with phases. Let us suppose that there is a discrete set of real characteristics, $S$, which is invariant under a 
suitable subgroup $\CG \subset {\rm Sp}(2\bar g, \IZ)$ of the full modular group. It follows that 
\be
\label{cftsum}
Z_{\Sigma, S}=\sum_{\mu,\nu \in S} \Bigl| Z_{\Sigma}(\mu,\nu)\Bigr|^2
\eeq
is invariant under ${\cal G}$. In a Calabi--Yau threefold $X$, the invariance subgroup of the modular group is the so-called {\it monodromy group} $\CM_X$, 
generated by the monodromies of the periods around the singularities in the moduli space of the spectral curve. 
It is natural to postulate that the nonperturbative structure of 
topological string theory on a local Calabi--Yau $X$ is given by a set $S$ of characteristics $\mu,\nu$ such that the sum (\ref{cftsum}) is invariant under 
the monodromy group $\CM_X$. We remark that the sum (\ref{cftsum}) is a ``diagonal" invariant, and it is conceivable that one can find non-diagonal invariants, as in the ADE classification of modular invariants for minimal models.  

As an example of this discussion, let us 
consider Seiberg--Witten theory for pure $\CN=2$ Yang--Mills theory with gauge group $SU(2)$, which can be 
regarded as a special limit of topological string theory (see \cite{lerche,klemm} for reviews of this model and its stringy origin). 
The spectral curve of this theory is the Seiberg--Witten curve $\Sigma_{\rm SW}$
\be
y^2=x\Bigl( x^2- u x +{1\over 4} \Bigr),
\eeq
 which is an elliptic curve with genus $\bar g=1$. 
The moduli space of this curve is the $u$-plane, with three singular points $u=\infty, 1,-1$. 
The monodromy group is $\Gamma^0(4)$, 
and it is generated by the monodromies around the singular points at $\infty$ and $u=1$, 
\be
M_{\infty} = \begin{pmatrix} -1 & 4 \\
0& -1\end{pmatrix}, \qquad M_1=\begin{pmatrix} 1 & 0\\
-1& 1\end{pmatrix}.
\eeq
Given the Seiberg--Witten curve, we can define nonperturbative partition functions for any characteristic $(\mu,\nu)$. It turns out that the characteristic 
$(0,1/2)$ is invariant under $\Gamma^0(4)$, therefore 
\be
\Bigl|Z_{\Sigma_{\rm SW}} (0,1/2)\Bigr|^2,
\eeq
is also invariant under the monodromy group. The theta function with this characteristic, $\vartheta_4$, 
appears naturally in Seiberg--Witten theory and in Donaldson--Witten 
theory in relation to the so-called blow-up function, see \cite{mmwhitham} for a review and a list of references.  

Another example is the manifold $Y_2$ in (\ref{localX2}). Its monodromy group is $\Gamma(2)$ \cite{abk}. 
In this case, one has from (\ref{charcs}) that $\mu=0$ and 
\be
\nu= {\pi \ri  \hat k \over 2} 
\eeq
and it depends on the value of the integer $\hat k=k+N$ mod $4$. 
For example, if $\hat k=4n +2$, $n \in \IZ$, 
one has $\mu=0, \nu=1/2$ which is also invariant under the monodromy group. 

The requirement of monodromy invariance of (\ref{cftsum}) might not be enough to completely fix the nonperturbative structure of the theory. It rather suggests that we should think of the nonperturbative partition functions $Z_{\Sigma}(\mu,\nu)$ 
as a set of conformal blocks associated to the mirror curve $\Sigma$. Finally, we note the natural appearance in this 
formalism of the modulus square of the nonperturbative partition function, which 
features in the OSV conjecture \cite{osv}. In our context, this is simply due to the requirement of modular invariance.

\sectiono{Integrability}

Here, we show that the non-perturbative partition function (\ref{exphatZTheta}), is a tau-function in the sense that it obeys Hirota equations.
We follow the same ideas as in \cite{eo}, which were applied there only for spectral curves $\Sigma=({\cal C},x,y)$ of genus $\bar{g}=0$, precisely because the $\Theta-$terms were missing in \cite{eo}.
So, here we complete the proof of \cite{eo} in the general case.

\medskip

For a spectral curve $\Sigma=({\cal C},x,y)$, let us write:
\beq
\tau(\Sigma) = \tau({\cal C},x,y)  = Z_\Sigma(\mu,\nu;\epsilon).
\eeq
We shall define the Baker-Akhiezer kernel through a Sato formula \cite{sato}:
\beq
K(z_1,z_2) = {1\over x(z_1)-x(z_2)}\,\,{\tau({\cal C},x,y+{1\over N}\,{\rd S_{z_1,z_2}\over \rd x})\over \tau({\cal C},x,y)}
\eeq
where $\rd S_{z_1,z_2}(z)$ is the 3rd kind differential having a simple pole at $z=z_1$ with residue $+1$, and a pole at $z=z_2$ with residue $-1$, and no other pole, and normalized on $A$-cycles, i.e. it is the integral of the Bergmann kernel:
\beq
\rd S_{z_1,z_2}(z) = \int_{z_2}^{z_1} B(z,z').
\eeq
In \cite{eo}, it is explained how to compute derivatives of $F_g$ with respect to any parameter of the spectral curve (see appendix \ref{appendixFg}), and in particular, if we consider the spectral curve $({\cal C},x,y+r\,{\rd S_{z_1,z_2}\over \rd x})$, we have:
\beq
{\partial \over \partial r} \,\, W_n^{(g)}(p_1,\dots,p_n) = \int_{z_2}^{z_1} W_{n+1}^{(g)}(p_1,\dots,p_n,z)
\eeq
and we may use this to compute the value of $\tau$ at $r=1/N$ by Taylor expansion:
\beq
\tau({\cal C},x,y+{1\over N}\,{\rd S_{z_1,z_2}\over \rd x}) = \sum_{k=0}^\infty {N^{-k}\over k!}\, \partial_r^k\, \tau({\cal C},x,y).
\eeq

This Taylor expansion can be performed more explicitly
\be
\label{BA}
\ba
&K(z_1,z_2)
= {\ee{N^2F_0+F_1}\over Z_\Sigma(\mu,\nu)}\,\,\,{\ee{-N\int_{z_2}^{z_1} ydx} \over E(z_1,z_2)\,\sqrt{\rd x(z_1)\rd x(z_2)}}\,\, \\
&  \quad \sum_{k} \sum_{n_i,l_i}\sum_{h_i>1-{l_i+n_i\over 2}} {N^{\sum_i (2-2h_i-l_i-n_i)}\over k! l_1!\,\dots\, l_k!\, n_1!\dots n_k!}\,\,\, \prod_{i=1}^k \left( \overbrace{\int_{z_2}^{z_1}\dots \int_{z_2}^{z_1}}^{n_i\,{\rm integrals}} \d_{\epsilon}^{l_i}\,W_{n_i}^{(h_i)} \right)  \,\, {\Theta_{\mu,\nu}^{(\sum_i l_i)}(NF'_0,\tau)}  \\
& \quad = {\ee{-N\int_{z_2}^{z_1} ydx} \over E(z_1,z_2)\,\sqrt{\rd x(z_1)\rd x(z_2)}}\,\, \\
& \Bigl\{ 1 + {1\over N}\Big(
 \int_{z_2}^{z_1} W_1^{(1)}  
 + {1\over 6} \int_{z_2}^{z_1} \int_{z_2}^{z_1} \int_{z_2}^{z_1} W_3^{(0)}
+ {1\over 2}  \int_{z_2}^{z_1} \int_{z_2}^{z_1}  {W_2^{(0)}}' {\Theta'\over \Theta} 
 + {1\over 2}  \int_{z_2}^{z_1} {W_1^{(0)}}'' {\Theta''\over \Theta}
 \Big) + \dots \Bigr\} \\
\ea
\eeq
where $E(z_1,z_2)$ is the prime form.
In this expression, each $\partial_\epsilon^{l_i} W_{n_i}^{(h_i)}$ is an abelian meromorphic form on ${\cal C}$, with no residue, and thus its integral from $z_2$ to $z_1$ is an abelian integral and defines a function of $z_1$ and $z_2$ only on the universal covering of ${\cal C}$.
If ${\cal C}$ is not simply connected, it is not obvious at all that expression (\ref{BA}) actually defines a function on ${\cal C}$.
However, thanks to background independence, this is the case: $K(z_1,z_2)$ is a function of $z_1,z_2\in {\cal C}$.
Indeed, if we move $z_1$ around a $B$-cycle, $\rd S_{z_1,z_2}$ gets shifted by the holomorphic differential $\om_i$:
\beq
\rd S_{z_1+B_i,z_2}  = \rd S_{z_1,z_2} + 2\ri\pi \om_i
\eeq
and thus moving $z_1$ around the ${\cal B}_i$ cycle, is equivalent to shifting $ydx$ by the holomorphic differential $2\ri\pi \om_i$, which is also equivalent to changing the filling fraction $\epsilon_i\to\epsilon_i+{1\over N}$. Since $\tau$ is independent of $\epsilon$, this means that 
\beq
\tau({\cal C},x,y+{1\over N}\,{\rd S_{z_1+B_i,z_2}\over \rd x})=\tau({\cal C},x,y+{1\over N}\,{\rd S_{z_1,z_2}+2\ri\pi \om_i\over \rd x}) = \tau({\cal C},x,y+{1\over N}\,{\rd S_{z_1,z_2}\over \rd x}) 
\eeq
and thus $K(z_1,z_2)$ is a well defined function of $z_1,z_2\in {\cal C}$.

\medskip

$K(z_1,z_2)$ has essential singularities when $z_1$ (resp. $z_2$) approaches a pole of $y\rd x$, and is such that:
\beq
\rd_{z_1}\,\ln{K(z_1,z_2)} \sim -N y(z_1)\rd x(z_1)
\eeq
$K(z_1,z_2)$ also has a simple pole at $z_1=z_2$:
\beq
K(z_1,z_2) \sim {1\over x(z_1)-x(z_2)}
\eeq
and those are the only singularities of $K$.

For any pair of spectral curves $\Sigma$ and $\td\Sigma$ over ${\cal C}$ we have the bilinear relation:
\beq
\Res_{z_2\to z_3}\, \rd x(z_2)\,\,K_\Sigma(z_1,z_2)\, K_{\td\Sigma}(z_2,z_3) = K_\Sigma(z_1,z_3).
\eeq
When written in terms of the moduli of $\Sigma$, this relation can be interpreted as the Hirota equation for the multicomponent KP hierarchy \cite{BookBDT}.

\section{Discussion}

Given an arbitrary spectral curve $\Sigma=({\cal C},x,y)$, and arbitrary characteristics $(\mu,\nu)$, we have defined a (non-perturbative) partition function (\ref{exphatZTheta}), and we have proved that it is both background independent and modular invariant, and it is perfectly holomorphic.
In other words, we have restored modular invariance not by breaking holomorphicity, but by breaking the perturbative expansion, i.e. by taking into account all instantons. 

One consequence of our analysis is that the holomorphic anomaly of the topological string might not be as fundamental as previously thought. If one thinks about the non-holomorphic 
dependence of the partition function as a way to restore modularity, one is tempted to 
say, in view of the results in this paper, that non-holomorphicity was the prize to pay in order to forget about nonperturbative corrections. This observation could also explain a long-standing puzzle for the large $N$ dualities of the topological string: the absence of a natural mechanism in the gauge theory side that might lead to non-holomorphic dependence on the 't Hooft couplings. In other words, there is no known, natural gauge theory dual of the holomorphic anomaly. According to our results, rather than mimicking the non-holomorphic way in which perturbative string theory achieves modularity, the gauge theory side restores 
the modularity properties of the full partition function 
by using non-perturbative effects, i.e. spacetime instantons.   

Another interesting outcome of our analysis is that it explains the similarity between the holomorphic anomaly equations and the heat equations for the theta functions 
that has been pointed out in \cite{witten} and recently elaborated in \cite{gnp,eone}. The nonperturbative partition function is simply the product of the perturbative 
partition function times a complicated theta function, as shown in (\ref{resumZ}). Since the product is modular, it follows that the perturbative piece has to transform precisely in 
a compensating way, therefore mirroring the transformation properties of theta functions. Notice however that the theta function involves $1/N$ corrections that depend on the 
perturbative free energies in a complicated way. It would be interesting to compare the nonperturbative paritition function studied in this paper to the holomorphic wavefunction 
introduced in \cite{gnp}, which was conjectured to be modular. 

The results presented in this paper can be improved in many aspects. The expression (\ref{exphatZTheta}) for the nonperturbative partition function 
is not completely satisfactory, since it is still defined by a $1/N$ expansion. 
It would be interesting to provide a more intrinsic formulation of $Z_{\Sigma}(\mu,\nu)$, maybe by using the integrability properties discussed in section 6. 
Each term in the expansion (\ref{exphatZTheta}) contains however a sum over all the multi-instantons. In this sense, the organization of the series is similar 
to what was done in \cite{cc} in the analysis of trans-series expansions: it is perturbative in $1/N$, but non-perturbative 
in $\re^{\pm N}$. 

It would be clearly important to understand the detailed implications of the nonperturbative completion that we are 
proposing. Since this completion involves multi-instanton corrections, it would be very interesting to relate them more precisely to the 
simplest cases studied in \cite{msw,mmnp,mswtwo}. These papers analyzed multi-instantons in the one-cut case, which should be regarded 
as a limiting case of the general situation considered in this paper (in fact, in \cite{mswtwo} the one-cut case was already regarded as a limit 
of the multi-cut case). One could use the multi-instanton information encoded in the nonperturbative partition 
function to understand the large order behavior of string 
perturbation theory and of the $1/N$ expansion of multi-cut matrix models, generalizing what was done in \cite{msw} in the one-cut case. 
Finally, it would be also interesting to relate the multi-instanton configurations encoded in $Z_\Sigma(\mu, \nu)$ to D-brane effects in topological 
string theory. 

On a more speculative note, we would like to point out that 
the nonperturbative partition function introduced in this paper incorporates 
a mechanism for recovering background independence which 
is similar in many respects to the mechanism suggested in 
\cite{maldacena} for restoring unitarity in some black hole backgrounds. In both cases one needs to sum over subleading saddle configurations which are invisible in the $1/N$ expansion. It is also interesting to notice that, when including nonperturbative corrections in the form of theta functions, the correlation functions of the matrix model display quasi-periodicity properties \cite{bde}, as required in theories without information loss (see the nice review in \cite{br} and references therein). Therefore, the 
mechanism to recover background independence studied in this paper might provide a useful toy model for the information paradox in the AdS/CFT correspondence. 
 
\section*{Acknowledgments}
We would like to thank Vincent Bouchard, Andrea Brini, 
Boris Dubrovin, Tamara Grava, Albrecht Klemm, Hirosi Ooguri, Sara Pasquetti, Ricardo Schiappa, Cumrum Vafa and Marlene Weiss 
for useful and fruitful discussions on this subject. B.E. would like to thank the Department of Mathematics of the 
University of Geneva for hospitality. The work of B.E. is partly supported by the Enigma European network MRT-CT-2004-5652, by the ANR project G\'eom\'etrie et int\'egrabilit\'e en physique math\'ematique ANR-05-BLAN-0029-01,  
by the European Science Foundation through the Misgam program,
by the Quebec government with the FQRNT. 
The work of M.M. is partially supported by FNS under the projects 200021-121605 and 200020-121675.

\appendix

\section{Symplectic invariants}
\label{appendixFg}
Here, we recall a few basic ingredients of the symplectic invariants $F_g$'s introduced in \cite{eo}.
\subsection{Spectral curve}

Consider a spectral curve $\Sigma=({\cal C},x,y)$, where ${\cal C}$ is a compact Riemann surface of genus $\bar{g}$ (not to be confused with the topological index $g$ of $F_g$), and $x$ and $y$ are two analytical functions on some open domain of ${\cal C}$.

The compact Riemann surface ${\cal C}$ is equipped with a symplectic basis of non-contractible cycles:
\beq
{\cal A}_i \cap {\cal B}_j = \delta_{i,j}
\eeq
and it comes with all classical tools of algebraic geometry (see \cite{Fay, Farkas}), in particular a prime form $E(z_1,z_2)$, which has a simple zero at $z_1=z_2$, and a Bergmann kernel $B(z_1,z_2)$, which has a double pole at $z_1=z_2$:
\beq
B(z_1,z_2) = {\rd f(z_1)\rd f(z_2)\over (f(z_1)-f(z_2))^2} + {\rm regular}
\eeq
in any local parametrization $f(z)$. $B$ is normalized on ${\cal A}$-cycles:
\beq
\oint_{{\cal A}_i} B = 0
\eeq
Explicitly we have ($c$ being an odd characteristics):
\beq
B(z_1,z_2) = \rd_{z_1}\rd_{z_2}\ln{\theta(z_1-z_2+c)}
\eeq
For example if ${\cal C}$ is a torus ${\mathbb C}/{\mathbb Z}+\tau{\mathbb Z}$, $B$ is the Weierstrass function $B(z_1,z_2) = \wp(z_1-z_2)\,dz_1\,dz_2$.

\medskip

The branch points are the points $a_i$ where $\rd x=0$. 
\beq
\rd x(a_i)=0
\eeq
We assume all branch points to be regular, i.e. $a_i$ is a simple zero of $\rd x$, and $\rd y(a_i)\neq 0$.
This means that. if $z$ is in the vicinity of $a_i$, there exists a unique $\bar{z}\neq z$ in the same vicinity of $a_i$ such that 
\beq
x(z)=x(\bar{z})
\eeq

\medskip

The recursion kernel $K$ is then:
\beq
K(z_1,z) = {\int^{\bar{z}}_z B(z_1,z')\over 2\,(y(z)-y(\bar{z}))\, \rd x(z)}
\eeq

\subsection{Correlators and symplectic invariants}

We define:
\beq
W_2^{(0)}(z_1,z_2) = B(z_1,z_2)
\eeq
and, if we denote $J=\{z_1,z_2,\dots,z_{n}\}$, we define recursively (on $2g+n$):
\bea
W_{n+1}^{(g)}(J,z_{n+1}) 
&=& \sum_i \Res_{z\to a_i}\, K(z_{n+1},z)\, \Big[ W_{n+2}^{(g-1)}(z,\bar{z},J) \cr
&& \qquad \quad + \sum_{h=0}^g\sum'_{I\subset J} W_{1+|I|}^{(h)}(z,I)W_{1+n-|I|}^{(g-h)}(\bar{z},J/I) \Big]
\eea
where $\sum'$ means that we exclude the terms $(h,I)=(0,\emptyset)$ and $(h,I)=(g,J)$.

\medskip

The symplectic invariants $F_g=W_0^{(g)}$ are defined by (if $g\geq 2$)
\beq
F_g = {1\over 2-2g}\, \sum_i \Res_{z\to a_i}\, W_1^{(g)}(z)\, \Phi(z)
\eeq
where $\rd \Phi=y \rd x$. 
For the definition of $F_1$ and $F_0$, we refer the reader to \cite{eo}.

\subsection{Derivatives}

The filling fraction is:
\beq
\epsilon_i = {1\over 2i\pi}\,\oint_{{\cal A}_i} y dx
\eeq
Changing $\epsilon_i$ is equivalent to changing $x$ and $y$ by:
\beq
{\partial\over \partial\epsilon_i}\, y \rd x 
= 2\ri\pi \om_i = \oint_{{\cal B}_i} B
\eeq
where $\om_j$ are the holomorphic abelian differentials on ${\cal C}$, normalized on ${\cal A}$-cycles:
\beq
\oint_{{\cal A}_i}\om_j =\delta_{i,j}
\eeq
Then \cite{eo} tells that:
\beq
{\partial \over \partial \epsilon_i}\, W_n^{(g)}(z_1,\dots,z_n) = \oint_{{\cal B}_i}\, W_{n+1}^{(g)}(z_1,\dots,z_n,z)
\eeq

\bigskip

Similarly, if we take a variation of $y\rd x$ such that:
\beq
{\partial\over \partial r}\, y \rd x 
= \rd S_{p_1,p_2} = \int_{p_2}^{p_1} B
\eeq
then \cite{eo} tells that:
\beq
{\partial \over \partial r}\, W_n^{(g)}(z_1,\dots,z_n) = \int_{p_2}^{p_1}\, W_{n+1}^{(g)}(z_1,\dots,z_n,z)
\eeq

\end{document}